\documentclass[aps,pre,twocolumn]{revtex4} 

\usepackage{setspace}
\usepackage{amsmath}
\usepackage{graphicx}
\usepackage{color}

\def\grad{\nabla}

\begin{document}

\title{Attractive asymmetric inclusions in elastic membranes under tension:\\ cluster phases and membrane invaginations}

\author{Sebastian Weitz$^{*}$, Nicolas Destainville$^{**}$}

\affiliation{Universit\'e de Toulouse; UPS; Laboratoire de
Physique Th\'eorique (IRSAMC); F-31062 Toulouse, France;\\
CNRS; LPT (IRSAMC); F-31062 Toulouse, France \\
{\rm *} Current address: Zentrum f\"ur Informationsdienste und Hochleistungsrechnen, Technische Universit\"at Dresden, Zellescher Weg 12, 01069 Dresden, Germany \\
{\rm **} {Corresponding author: \tt nicolas.destainville@irsamc.ups-tlse.fr}
}
\date{\today}

\begin{abstract}
Up-down asymmetric inclusions impose a local, spontaneous curvature to an elastic membrane. When several of them are inserted in a same membrane, they feel effective forces mediated by the membrane, both of elastic and entropic nature. Following an approach initiated by Dommersnes and Fournier in the vanishing tension case [Eur. Phys. J. B {\bf 12}, 9 (1999)], and also using a pseudo-analytical micellization theory, we derive the statistical mechanics of asymmetric inclusion assemblies when they are also subject to an additional short-range, attractive interaction. Our main conclusion is that generically, when the membrane is under tension, these inclusions live in small clusters at equilibrium, leading to local membrane invaginations. We also propose a novel curvature-induced demixing mechanism: when inclusions imposing local curvatures of opposite sign coexist, they tend to demix in distinct clusters under realistic conditions. This work has potential implications in the context of the thermodynamics of proteins embedded in biological lipid bilayers. 

\end{abstract}

\maketitle

\section{Introduction}

Having in mind the elucidation of the physical laws governing assemblies of proteins embedded in lipidic biomembranes~\cite{Phillips09}, the statistical mechanics of membrane inclusions has been the focus of active and plethoric research in the two past decades~\cite{Dan93,Goulian93,Park96,Golestanian96,Fournier97,Fournier98,Weikl98,Domm99,Sear99,Kim99, Marchenko02, Fournier03,Bohinc03,Sieber,Periole08,Bibi08,Foret08, Gurry09,Weiss09,Meyer10,Bibi10,Lin11,Meilhac11,Markova12,Rob13}. Several complementary theoretical and numerical developments have been proposed to study these soft matter systems, among which mechanical approaches in the zero-temperature limit~\cite{Bohinc03,Goulian93,Fournier97,Kim99}, field-theorectical studies taking explicitly membrane fluctuations into account~\cite{Goulian93,Domm99,Fournier03,Lin11}, Molecular Dynamics~\cite{Reynwar07,Periole08,Weiss09,Meyer10} or mesoscopic-scale Monte Carlo simulations~\cite{Fournier03,Bibi08,Meilhac11}. However, at this stage, no consensual view of the dynamical membrane organization exists that is fully recognized by both biologists and physicists. In particular, the physical mechanism by which membrane proteins congregate in small clusters of tens to hundreds of entities, as observed experimentally~\cite{NSOM,Lenne09,Lang10,Bogaart11,Schreiber12}, remains to be elucidated. 

The present statistical mechanics work participates to an approach where finiteness of clusters is attributed to an equilibrium argument based upon the competition between short-range attractive forces that favor inclusion condensation and a longer-range, weaker repulsion which forbids a complete condensation because too large clusters become unstable~\cite{Sieber,Gurry09,Evans03}. The result is a stable (or metastable) ``cluster phase''~\cite{Bibi08,Foret08}. This is yet another example where the competition between attraction and repulsion leads to the existence of patterns at equilibrium~\cite{Seul95}. From a biological perspective, this approach does not exclude that out-of-equilibrium effects leading to dynamic cluster remodeling~\cite{Foret05} can modify cluster-size distributions as compared to equilibrium ones. However it assumes that an equilibrium argument is required because clusters survive even in the absence of active processes, e.g. in inactive membrane sheets~\cite{Halemani10,Schreiber12}.

\begin{figure}[ht]
\begin{center}
\includegraphics*[width=7cm]{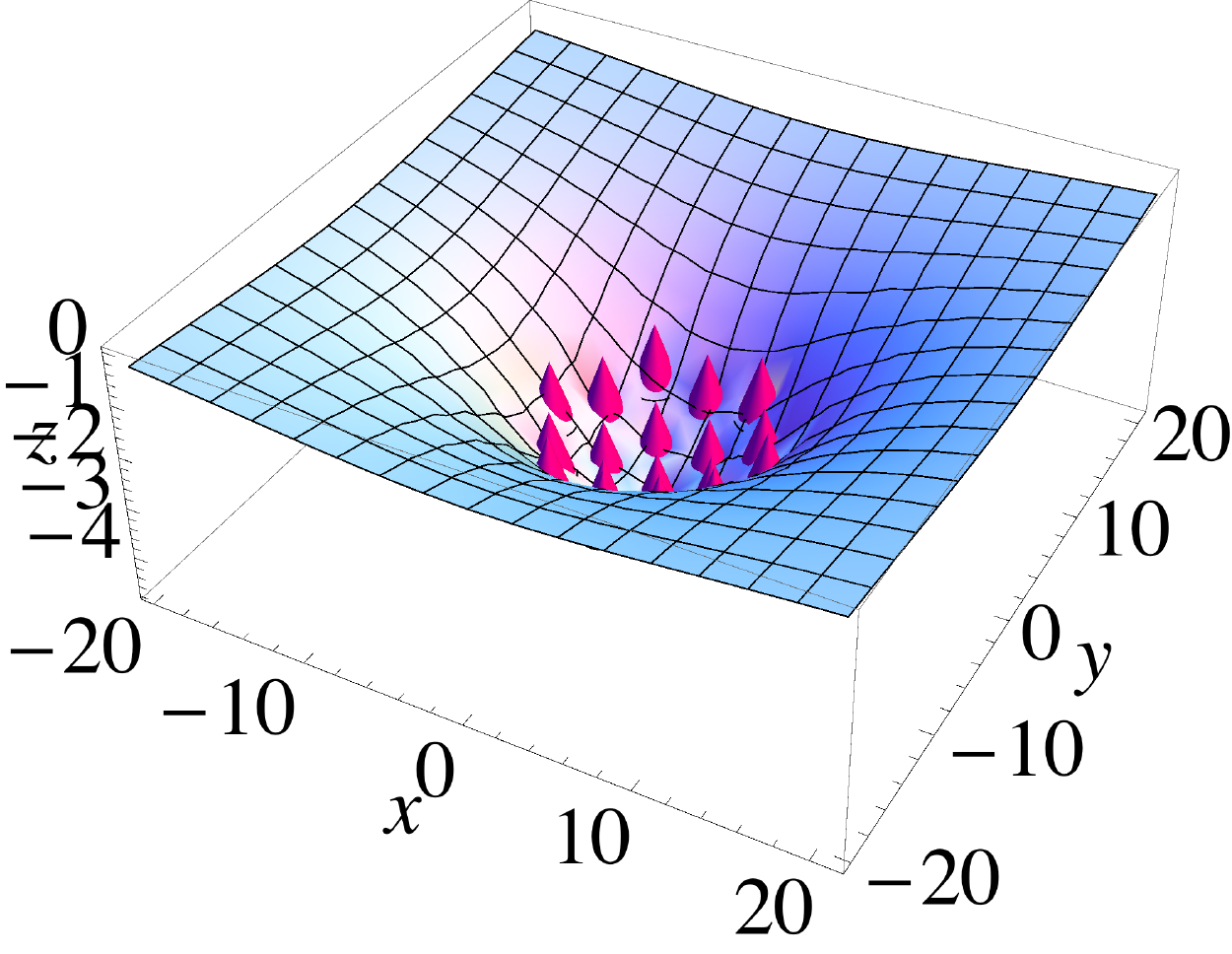}
\end{center}
\caption{Example of membrane invagination induced by a cluster of $n=19$ identical asymetric inclusions (pink cones). Each inclusion imposes a local curvature and the elastic membrane minimizes its free energy by budding in the $z$ direction. 
\label{profile1}}
\end{figure}

In this approach, the repulsion between inclusions is mediated by the elastic membrane, and comes from the more or less pronounced up-down asymmetric character of the membrane deformation imposed by the inclusions~\cite{Bibi08}. When inclusions are aggregated in a small cluster, the membrane mechanical response leads to its invagination~\cite{Fournier03}, as illustrated in Figure~\ref{profile1}, and we shall see that the resulting equilibrium free energy grows faster than the cluster size: it is non-extensive. This leads to the instability of too large clusters~\cite{Foret08} and the equilibrium configuration is a cluster phase, where small clusters co-exist at equilibrium with a gas of monomers. The goal of the present work is to analyze this cluster phase mechanism in greater detail by thoroughly addressing some important, but still unexplored questions, notably the exact role played by many-body forces in the repulsive interaction, the role of membrane tension, and the precise behavior of asymmetric inclusions with opposite orientations. 

In the biological context, a light up-down asymmetry can result from the crystallographic shape of membrane proteins, which can be conical and/or wedge-shaped~\cite{Doyle98,Bass02,Park08,Manglik12},Ê or from more subtle effects such as up-down asymetric electrostatic distributions or even leaflet asymmetry~\cite{reviewSeifert}. The way proteins insert into the membrane can also be inferred from numerical approaches as well as molecular or biochemical techniques~\cite{vanGeest00}. It is precisely the elastic repulsion ensuing from the up-down asymmetry that will be explored in details in the present work, thus giving a more solid theoretical foundation to the equilibrium scenario discussed above.

The present works adopts a field-theoretical formulation~\cite{Goulian93,Domm99,Fournier03,Lin11} because it efficiently and elegantly provides exact results without requiring too tedious calculations. It is pseudo-analytical: after integrating Gaussian functional integrals, free-energies can be expressed in terms of inverses and determinants of large matrices. Therefore a light numerical work is required at calculation end. However, no intensive numerical work such as Monte-Carlo sampling is required.

We make some simplifying assumptions throughout this paper: we consider that the membrane contains a single inclusion species, even though asymmetric inclusions will possibly have two opposite orientations. Furthermore, we assume the membrane to have a finite tension $\sigma> 0$, because this is the case of biophysical relevance. It has already been shown in related contexts that tension can play a determinant role~\cite{Weikl98,Dean06}. The vanishing tension limit will nevertheless be thoroughly discussed. 

The paper is organized as follows. After summarizing in Section~\ref{theo} anterior theoretical works that will be useful in the present context, we first tackle the simple case of two and three-body interactions (Section~\ref{2:3}). We then switch to the general case under interest, where a large number of inclusions interact, and analyze the requirements for the existence of cluster phases (Section~\ref{Micel}). Section~\ref{opposite} is devoted to the study of systems of inclusions imposing curvatures of opposite orientations. We conclude and give perspectives in Section~\ref{cl}.
In this paper, we use reduced units: when they are dimensionless, energies are implicitly in units of the thermal energy $k_{\rm B}T$ ($k_{\rm B}$ is Boltzmann's constant), and lengths are in units of the inclusion radius $a$. 

\section{Summary of the theoretical framework}
\label{theo}

We first display the statistical mechanics of an assembly of $P$ inclusions embedded in an elastic sheet of curvature elastic modulus $\kappa>0$ and tension $\sigma>0$, and situated at fixed positions $\mathbf{r}_p \in \mathbf{R}^2$. Following the same procedure as Fournier and Dommersnes~\cite{Domm99,Fournier03}, we extend their work in the non-vanishing tension case. This section also recalls the approximate micellization theory of Reference~\cite{Foret08} and explains how it enables one to compute cluster-size distributions (when a cluster phase does exist) at equilibrium. 

\subsection{Exact membrane/inclusions free energy~\cite{Domm99,Fournier03}}

To calculate the free energy of the membrane/inclusions system, we first model the continuous lipidic bilayer by Helfrich's free energy~\cite{Helfrich73}
\begin{eqnarray}
F[u]=
\frac{\sigma}{2}
\int_{\mathcal S} d^2\mathbf{r}
\left[
(\nabla u)^2
+
\lambda^2 (\Delta u)^2
\right],
\label{eq:F-partielle}
\end{eqnarray}
where $u(\mathbf{r})$ is the membrane height function above a reference plane and the distance $\lambda\equiv\sqrt{\kappa/\sigma}$ will be shown below to be a typical interaction range and to play an important role. Note that, in principle, Helfrich's approach is only valid in the limit where the derivatives of $u$ involved in $F[u]$ are sufficiently small. 

Following References~\cite{Domm99,Fournier03}, a membrane inclusion at position $\mathbf{r}_p \in \textbf{R}^2$ is assumed to locally impose the membrane curvature tensor $\grad\grad u|_{\mathbf{r}_p}$. The imposed value is denoted by  $Q_p$, a symmetric $2\times 2$ real matrix. Assuming that there are $P$ such inclusions in a membrane patch, the canonical partition function is given by the Gaussian functional integral~\cite{Domm99,Fournier03}
\begin{eqnarray}
\label{eq:Z}
Z_P=\int {\mathcal D}u \ 
\prod_{p=1}^{P}
\delta(\grad\grad u|_{\mathbf{r}_p}-Q_{p})
e^{-\beta F[u]}
\end{eqnarray}
where $\beta \equiv 1/(k_{\rm B}T)$ is the inverse temperature. The symmetric tensor can always be diagonalized in normal coordinates, $Q_p=P^{-1}Q_p^{\rm diag}P$, with 
\begin{equation}
Q_p^{\rm diag}=\left(
\begin{array}{cc}
C_p+J_p & 0 \\
0 & C_p-J_p
\end{array}
\right).
\label{diagform}
\end{equation}
This is the most general case. When $J_p=0$, the inclusion is said to be isotropic. This is the case we focus on in this work. When both $C_p$ and $J_p$ vanish, it is said to be (up-down) symmetric. Two inclusions with $C_p$ values of different signs will be said below to be of opposite orientation (or ``head-to-tail'').

The Helfrich propagator reads
\begin{equation}
G(r\equiv|\mathbf{r}|)=-\frac{1}{2\pi\sigma}
\left[
K_0\left(\frac{r}{\lambda}\right)
+
\ln \left(\frac{r}{\lambda}\right)
\right]
\label{propag}
\end{equation}
where the $K_\nu$ (here and below) are the modified Bessel functions of the second kind. Note that because only derivatives of $G$ will be used below, $G$ can be defined up to a constant. 

After subtraction of the free energy of the membrane without any inclusion, the system free energy becomes~\cite{Domm99,Fournier03}
\begin{equation}
F_P\equiv -k_{\rm B}T \ln Z_P
=
\frac{k_{\rm B}T}{2}\ln\det M
+
\frac{1}{2} \mathbf{Q}^T M^{-1}\mathbf{Q},
\label{eq:F}
\end{equation}
where: (i) $\mathbf{Q}$ is a vector of $\mathbf{R}^{3P}$, $\mathbf{Q}=(\ldots,a_p,b_p,c_p,\ldots)$, where $a_p$ and $b_p$ are the two diagonal elements of $Q_p$ (before its diagonalization) and $c_p$ is the off-diagonal one ($Q_p$ is symmetric); and (ii) $M$ is a $3P\times3P$ block matrix defined as follows: each block $m_{pq}$ is a $3\times3$ real symmetric matrix. For $p\neq q$, $m_{pq}=\mathbf{D}\mathbf{D}^TG(\mathbf{r}_p - \mathbf{r}_q)$ where $\mathbf{D}$ is the operator-vector of coordinates $(\partial^2_x,\partial^2_y,\partial_x\partial_y)$. Diagonal blocks $m_{pp}$ deserve a special attention because the previous definition of $m_{pq}$ displays an Ultra-Violet (UV)  divergence when $p=q$ and require the introduction of an UV cutoff $q_{\rm max}\equiv 1/r_0$~\cite{Fournier03}. Then 
\begin{eqnarray}
m_{pp}=
\frac{q_{\rm max}^2}{32\pi\kappa}
\left[1-\frac{1}{\lambda^2q_{\rm max}^2}
\ln\left(1+\lambda^2 q_{\rm max}^2\right)
\right]
\begin{pmatrix}
3&1&0\\
1&3&0\\
0&0&1\\
\end{pmatrix}.
\label{diag}
\end{eqnarray}

Introducing an UV cutoff is necessary because when using a point-particle description ~\cite{Domm99,Marchenko02,Fournier03}, one forgets the actual size of the particles. The choice of the cutoff value $r_0$ is delicate~\cite{Yolcu12}. Here, following Reference~\cite{Fournier03}, we chose $r_0$ to be proportional to the inclusion size $a$ (the only available short length-scale): in order to recover the good interaction potentials at long range as compared to calculations with finite-size inclusions~\cite{Goulian93,Fournier97,Weikl98}, one has to set $r_0=a/2$. Note however that there exist less ambigous, more sophisticated approaches appealing to renormalization arguments~\cite{Lin11,Yolcu12}. 

In equation~(\ref{eq:F}), the first r.h.s. term is the so-called Casimir interaction. It is purely entropic in nature and will be denoted by $F^{\rm C}$. The second term will be called the purely elastic contribution ($F^{\rm el} $), because $F_P=F^{\rm C}+F^{\rm el}$ tends to this purely elastic part in the zero-temperature limit where membrane fluctuations~\cite{Safran} and entropy vanish. This exact form, equation~(\ref{eq:F}), will be useful in sections \ref{Micel} and \ref{opposite} below. In the vanishing tension limit ($\lambda\to\infty$), one recovers Dommersnes and Fournier's results~\cite{Domm99,Fournier03}. 

Finally, the average membrane shape resulting from the presence of the inclusions at positions $\mathbf{r}_p$ is given by~\cite{Domm99,Fournier03}:  
\begin{equation}
\langle u(x,y) \rangle = \mathbf{c}^T M^{-1}\mathbf{Q},
\label{av:shape}
\end{equation}
where $\mathbf{c}$ is again a vector of $\mathbf{R}^{3P}$, 
$\mathbf{c}=(\ldots,\partial_x^2 G(\mathbf{r}-\mathbf{r}_p),\partial_y^2 G(\mathbf{r}-\mathbf{r}_p),\partial_x \partial_y G(\mathbf{r}-\mathbf{r}_p),\ldots)$,
with $\mathbf{r}=(x,y)$. Figure~\ref{profile1} provided an example.

\subsection{Approximate micellization theory~\cite{Foret08}}
\label{MicelTheo}

The previous exact approach enables one to calculate the free energy of any cluster of size $n$ (i.e. containing $n$ inclusions), denoted by $F(n)$, provided that a prescription for the short-range attraction is also given. Sections \ref{Micel} and \ref{opposite} below will be dedicated to this issue. Once these free energies $F(n)$ are known, we can apply the approximate scheme of Reference~\cite{Foret08} as follows. We assume that $F(n)$ can be written in the (generalized) droplet form taking into account both the short-range attraction and the longer-range repulsion:
\begin{equation}
F(n)=-f_0(n-1)+\rho_0 \sqrt{n-1} + \chi (n-1)^\alpha,
\label{droplet:form}
\end{equation}
where $f_0,\rho_0,\chi>0$ and $\alpha>1$~\footnote{We choose to expand the cluster free energies $F(n)$ in function of $n-1$~\cite{Foret08} (and not of $n$) so that it vanishes for $n=1$ (i.e. for monomers), as required, since we have subtracted the inclusion self-energies from $F$.}. The cluster-bulk free energy per particle $f_0$ includes the translational entropy of each particle in the cluster~\cite{Foret08}. If the average coordination number (the number of neighbors of each inclusion in a cluster bulk) is $c$ and if the binding energy between two neighbors in a cluster is $\varepsilon_a$, then $f_0 \approx c \varepsilon_a/2$. The parameter $\rho_0$ is a line tension, which expresses the fact that particles on the cluster boundary have less neighbors than bulk ones. The value of $\rho_0$ can be estimated by calculating the number of boundary particles in a circular cluster and writing that each boundary inclusion has typically twice less neighbors than a bulk particle. In the case where the particles are arranged on a triangular lattice ($c=6$), a simple calculation leads to $\rho_0 \approx  (\sqrt{\pi}3^{1/4}/2) f_0 \approx f_0$. We denote by $F^{\rm s.r.}(n)=-f_0(n-1)+\rho_0 \sqrt{n-1}$ this ``bare'' short-range contribution. However we shall see below that the ``bare'' values of $f_0$ and $\rho_0$, resulting from the short-range forces alone, are likely to be renormalized by the interactions propagated by the elastic membrane. Finally, the last term in Equation~(\ref{droplet:form}) comes from the long-range repulsion and is responsible for the instability of too large clusters. It will be discussed in great detail below.

Next, we switch to the grand-canonical ensemble to impose the total number of inclusions, denoted by $N$, via their chemical potential $\mu$. More precisely, if the system area is denoted by A and if $D=2a$ is the particle diameter, we impose the adimensional surface fraction $\phi \equiv ND^2/A$, which sets $\mu$ as in~\cite{Foret08}: if the number of $n$-clusters is denoted by $N_n$ and their mean surface fraction is $c(n)\equiv\langle N_n \rangle D^2/A$, then $c(n)=e^{\mu n-F(n)}$~\cite{Safran} and $\mu$ is obtained by solving the equation $\phi=\sum_{n=1}^{\infty} n c(n)$.
Once $\mu$ is known, the cluster-size distribution $c(n)$ follows. 

This approach contains several approximations, notably the droplet form~(\ref{droplet:form}), which, in principle, is only valid for large values of $n$, as well as the fact that  interactions between clusters are ignored. Inter-cluster interactions are indeed negligible, at least at low $\phi$, because of the rapid decay of the interactions at distances larger than $\lambda$, as illustrated in the next section.

\section{Two- and three-body interactions}
\label{2:3}
Before focusing on clusters, we first address the calculation of two- and three-body forces, where we bring some ameliorations as compared to anterior work by Weikl, Kozlov and Helfrich~\cite{Weikl98}. 
\begin{figure}[ht]
\begin{center}
\includegraphics*[width=7cm]{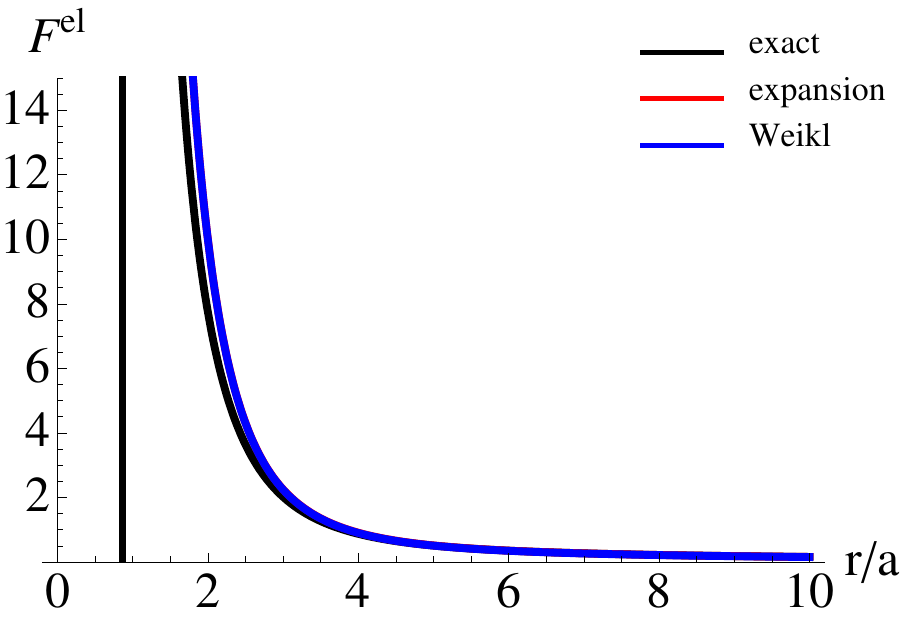}
\includegraphics*[width=7cm]{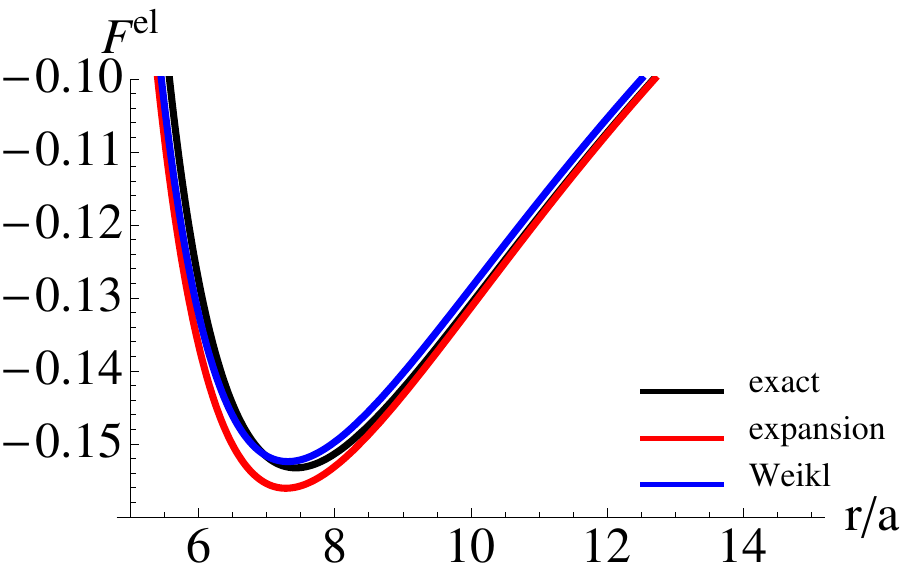}
\end{center}
\caption{Exact two-body purely elastic interaction $F^{\rm el}$ in function of the inclusion separation $r$ [second r.h.s. term in equation~(\ref{eq:F}) with $N=2$ inclusions, in black]; Our large $r$ expansion [equations~(\ref{F22}) and (\ref{F42}) with $N=2$ inclusions, in red, superimposed with the blue curve in the top panel]; Weikl et al.'s expansion~\cite{Weikl98} [in blue]. The curvature modulus is $\kappa=153$, the tension is $\sigma=1$, the contact angles are $\gamma_1=\pm \gamma_2=0.2$~rad. Top: identical inclusions ($\gamma_1=\gamma_2$); Bottom: inclusions with opposite orientation (i.e. head-to-tail inclusions, $\gamma_1=- \gamma_2$).
\label{profilApp}}
\end{figure}

\begin{figure}[ht]
\begin{center}
\includegraphics*[width=7cm]{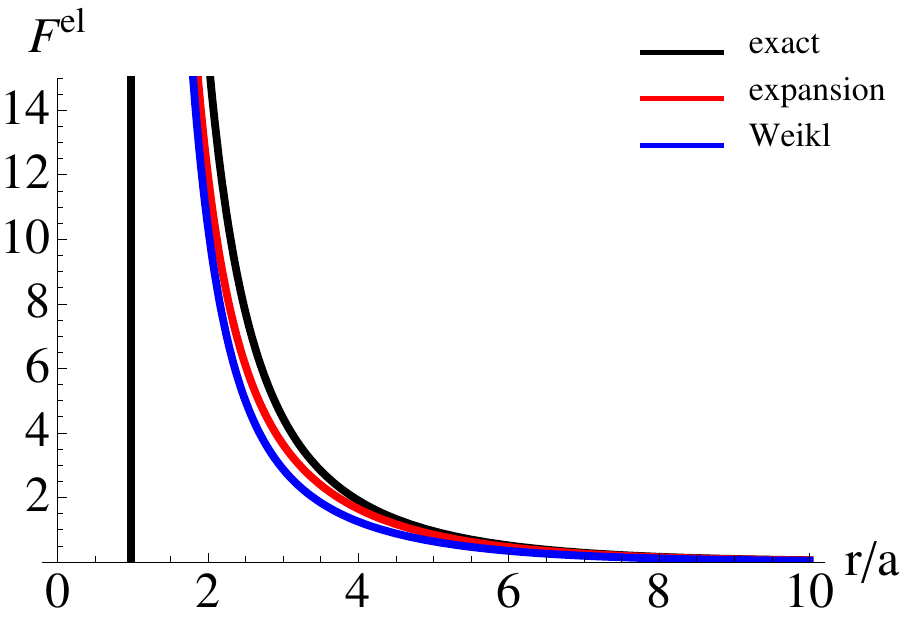}
\includegraphics*[width=7cm]{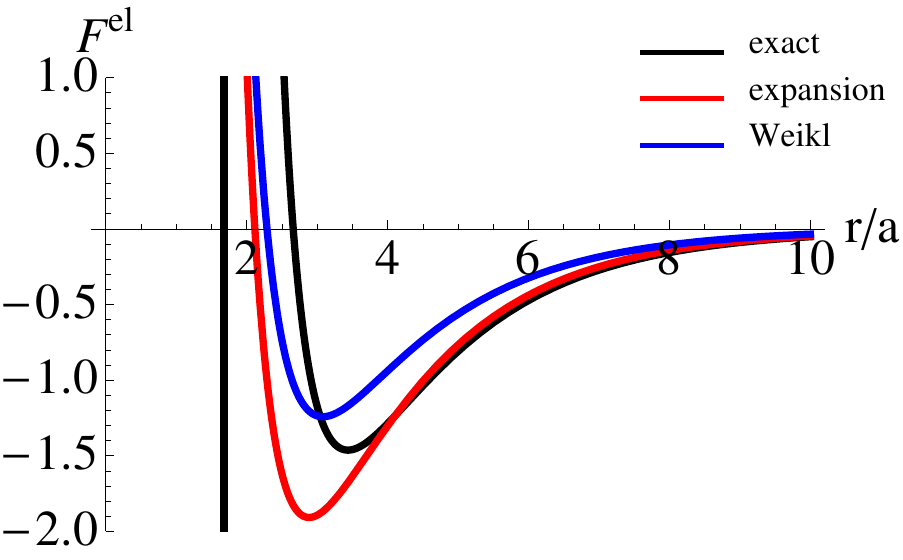}
\end{center}
\caption{Same as Figure~\ref{profilApp} except that the tension is now very high, $\sigma=38$.
\label{profilApp2}}
\end{figure}

In order to write an expansion of $F$, we first note that owing to equations~(\ref{propag}) and (\ref{diag}) and to the definition $\lambda^2=\kappa/\sigma$, the block sub-matrices of $M$ can be written as $1/(\kappa\lambda^2)$ times a dimensionless function of $q_{\rm max} \lambda$ where $q_{\rm max} = 1/r_0 \propto 1/a$. Thus the free energy is a function of $a/\lambda$; It is also a function of $r/\lambda$ owing to the expression of the propagator $G$ (see the precise expressions of these expansions below). In practice, we are interested in the limit $a \ll \lambda$ of small inclusions, thus it is relevant to expand $F$ in powers of $a/\lambda$. By contrast, $r/\lambda$ enters Bessel functions that decay exponentially at large arguments, logarithms or power-laws. Furthermore, values of $r$ either smaller or larger than $\lambda$ are of interest. Thus we shall not write expansions in function of $r/\lambda$. In addition, we introduce the dimensionless quantities $\gamma_k=aC_k$ that can be assimilated to inclusion contact angles when comparing our results to anterior ones~\cite{Goulian93,Fournier97,Weikl98}. If an isotropic protein is modeled by a conic inclusion, the contact angle $\gamma$ is, by definition, half the cone aperture. 

To begin with, we explore the purely elastic contribution $F^{\rm el}$. The expansion of $F^{\rm el}$ in powers of $a/\lambda$ is obtained from the second term in the r.h.s of equation~(\ref{eq:F}) and reads:
\begin{equation}
F^{\rm el}=F^{\rm el}_0+F^{\rm el}_{22}+F^{\rm el}_{42}+F^{\rm el}_{43}+\mathcal{O}\left[(a/\lambda)^6 \right].
\label{eq:Fel-DLa6}
\end{equation}
In this expression, $F^{\rm el}_0$ is  a self-energy, independant of
the inter-inclusion distances, and will not be considered further;
$F^{\rm el}_{22}$ and $F^{\rm el}_{42}$ are the two dominant two-body interactions
[where we introduce the simplified notation $K_\nu^{(ij)}\equiv K_\nu(r_{ij}/\lambda)$]:
\begin{eqnarray}
F^{\rm el}_{22}&=&2\pi\kappa 
\frac{a^2}{\lambda^2}\sum_{i>j}\gamma_i \gamma_j K_0^{(ij)}
\label{F22}
\end{eqnarray}
\begin{eqnarray}
F^{\rm el}_{42}&=&\frac{\pi}{2}\kappa 
\frac{a^4}{\lambda^4}\sum_{i>j} \Bigg\{
(\gamma_i^2+\gamma_j^2)\left[{K_0^{(ij)}}^2+2{K_2^{(ij)}}^2\right] 
\nonumber \\
&&+4\ln\frac{2\lambda}{a}\gamma_i \gamma_j K_0^{(ij)}\Bigg\},
\label{F42}
\end{eqnarray}
for isotropic inclusions. Finally, $F^{\rm el}_{43}$ is the dominant three-body potential: 
\begin{eqnarray}
F^{\rm el}_{43}&=&\pi\kappa 
\frac{a^4}{\lambda^4}\sum_{i,k>j}' \gamma_j \gamma_k 
\left[ K_0^{(ij)}K_0^{(ik)} \right.  \nonumber \\
&&+\left.2\cos(2\alpha^i_{jk})K_2^{(ij)} K_2^{(ik)}\right].
\label{F43}
\end{eqnarray}
where the primed sum indicates that the three indices are different~\cite{Domm99}. As in this Reference, $\alpha^i_{jk}$ is the angle between the line joining inclusions $i$ and $j$ and the line joining inclusions $i$ and $k$. 


We again checked that we recover known results in the vanishing-tension limit~\cite{Domm99}\footnote{Our calculation for $J_p\neq 0$ in the vanishing-tension limit leads to a slightly different result compared to Reference ~\cite{Domm99}: the $\cos(2\theta_{pn}-2\theta_{np})$ term in equation (10) of ~\cite{Domm99} has to be replaced by $\cos(2\theta_{pn}+2\theta_{np})$. Our result is supported be the fact that the free energy has to be minimized for $\theta_{np}\approx -\theta_{pn}$ to be in agreement with the ``egg-carton" structure of the membrane obtained by Monte Carlo simulations (using the whole $N$-body interactions) in Figure 2a of ~\cite{Domm99}.}. Our two-body expansion can now be compared to Weikl, Kozlov and Helfrich's one~\cite{Weikl98}. Comparing their Equation~(23) to our Equation~(\ref{F42}), one remarks that in the $(a/\lambda)^4$-order term of the expansion, they only found the sub-dominant ${K_2^{(ij)}}^2$ sub-term. As illustrated in Figures~\ref{profilApp} and \ref{profilApp2}, this leads to differences between our expansion and theirs, especially at large inclusion separations. These differences are comparable for identical and head-to-tail inclusions: at low tension of biological interest, both our expansion and Weikl et al.'s one are very good, whereas at larger tensions, the differences are more pronounced. In every case, at short separations $r$, both expansions fail to reproduce the correct behavior, whereas at large separations $r \gtrsim \lambda$, our expansion is better than Weikl et al.'s one. We have explored a wide range of parameters (as discussed in the next section \ref{REF}) and our conclusions remain the same. 

Finally, the expansion of the Casimir interaction $F^{\rm C}$ is given in Appendix~\ref{purcasimir}. It does not depend on the curvatures, as evidenced in Equation~(\ref{eq:F}), and will be shown to play a marginal role because it does not contribute to the long-range term $\chi (n-1)^\alpha$ in Equation~(\ref{droplet:form}). Note that this expansion contains algebraically-decaying terms (in $1/r^8$), as already pointed up in Reference~\cite{Lin11}. However this exponent 8 is too high to lead to any non-extensivity~\cite{Foret08}.

\section{Clusters in the micellization theory framework}
\label{Micel}

Beyond two- and three-body interactions, $N$-body forces are likely to play a significant role inside clusters where inclusions are densely packed. To explore the role of these forces, it is not reasonable to compute and introduce successively 4-body, 5-body, etc, terms in the calculation of the energy, because of their growing complexity. By contrast, the formalism introduced in section~\ref{theo} gives access to the complete, exact long-range forces mediated by the membrane for a cluster of any size provided that one is able to invert the $3P \times 3P$ matrix $M$. In practice, systems containing $P\sim 100$ particles are easily accessible, as detailed below.

However, at short, nanometric range, effective interactions are not as well controlled because (i) inclusions are expected to interact but, due the lack of exact knowledge of these short-range forces, we model them in a very basic, two-body fashion (see discussions in References~\cite{Foret08,Meilhac11} and references therein; For example, this point of view neglects some effects such as the many-body character of hydrophobic mismatch forces~\cite{Brannigan07}); (ii) a continuous description of biomembranes cannot be anticipated to be exact at length scales on the order of the size of its elementary constituents, namely lipids in the present case. Therefore the present study contains some degree of approximation (as discussed, e.g., in Reference~\cite{Fournier03}). Numerical simulations can give more insight at short range~\cite{Periole08,Weiss09} even though they also rely on approximations and assumptions, and  have their own limitations.

\subsection{Reference parameter sets}
\label{REF}

Before going on, we need to introduce some reference parameter sets, which will be useful in the following. They intend to be realistic in the context of cell membranes, as motivated in the introduction. In all cases, the inclusion radius is fixed to $a=2$~nm, a typical membrane protein radius~\cite{Bibi08}. Unless explicitly specified, the contact angles $\gamma_k$ are set to $\pm0.2$~rad~\cite{Fournier03}, i.e. to $\pm 11.5^\circ$.
\begin{itemize}
\item $\mathcal{R}$ parameter set: The curvature modulus is $\kappa=50$~$k_{\rm B}T$, a commonly accepted value for lipidic bilayers: depending on the membrane composition, $\kappa$ has been measured to be on the order of 10 to 100~$k_{\rm B}T$ or even more~\cite{Daoud99,Mouritsen}. The membrane tension is chosen to be $\sigma=1$ in reduced units, i.e. $k_{\rm B}T/a^2 \simeq 10^{-3}$~J/m$^2$, so that the interaction range is $\lambda=7.1$, i.e. 14.2~nm. This tension is one order of magnitude lower than typical lipidic membranes lysis tensions~\cite{Evans76,Sheetz96}. It is intentionally large because we want to unambiguously address tension effects. It will have to be lowered below when adressing moderate or weak membrane tensions of cell-biology interest, on the order of $10^{-4}$~J/m$^2$ or even lower~\cite{Needham92,Sheetz96}. This corresponds to $\sigma \lesssim 0.1$ in reduced units. The depth of the short-range attraction well is $\varepsilon_a=4$ $k_{\rm B}T$~\cite{Bibi08,Weiss09}. Finally, in clusters, particles are arranged on a compact triangular lattice~\cite{Bibi08} of spacing $l=2a=D$, the particle diameter. The coordination number is thus $c=6$, which sets $f_0 \approx \rho_0 \approx 12$~$k_{\rm B}T$ (see Section~\ref{MicelTheo}). 

\item $\mathcal{R}'$ parameter set: Same as $\mathcal{R}$ except that the hexagonal lattice spacing becomes $l=3.5a$ to avoid some spurious singularities of the free energy, as discussed later. This can for instance be attributed to dilation effects that move constituents of a cluster apart, or, in the biophysical context, to enhanced hard-core repulsion due to protruding extra-membrane regions of transmembrane proteins, larger than their intra-membrane hydrophobic core~\cite{Sieber}. The elastic modulus $\kappa$ is also rescaled to $\kappa=153$ $k_{\rm B}T$ so that the ratio $\lambda/l$ remains unchanged as compared to $\mathcal{R}$ (see Section~\ref{manybody}). If dilation effects are appealed to, random deviations from the regular lattice due to thermal agitation should in principle also be taken into account. This possibility will be tackled at the end of Section~\ref{manybody} below. Figure~\ref{profile1} was calculated using this $\mathcal{R}'$ parameter set. 

\item $\mathcal{R}''(\lambda)$ parameter set: Same as $\mathcal{R}'$ except that the range $\lambda=\sqrt{\kappa/\sigma}$ is modulated by changing the value of the tension $\sigma$, at constant $\kappa=153$ $k_{\rm B}T$. This parameter set will be useful when addressing lower tension regimes. Note that $\mathcal{R}'=\mathcal{R}''(\lambda=\sqrt{153}\simeq12.4)$.
\end{itemize}

In all cases below, when speaking of a circular cluster of size $n$, or circular $n$-cluster, one must understand all the particles of the triangular lattice lying inside a circle of random center and of adequate radius, so that it exactly embraces $n$ particles, as exemplified in Figure~\ref{profile1}. All relevant observables are then averaged over $R\gg1$ realizations, one realization corresponding to the uniform random choice of the circle center position in the elementary lattice cell. This procedure smoothes spurious effects due to irregular boundaries, in particular for small $n$. This is the only non-analytical part of the free energy calculation.

\subsection{Singularities of the free energy at short lattice spacing}
\label{singul}

The pair potentials displayed in Figures~\ref{profilApp} and \ref{profilApp2} present a singularity, but at a distance shorter than the particle hard-core diameter, $D=2a$.  This singularity is thus physically irrelevant. However, when the number $n$ of particles in a dense cluster increases, this issue becomes more serious because we observe the singularity to appear at lattice spacings $l$  larger than $D$. The occurrence of such singularities stresses the fact that the elastic theory used in the present work has intrinsic physical limitations, already foreseen when we were led to introduce the cutoff $q_{\rm max}$ above. Another manifestation of this issue is as follows. If $l$ is too short, when drawing the membrane shape as in Figure~\ref{profile1}, one observes  that the invagination is inverted, with a curvature sign opposed to the one imposed by the inclusions (not shown). This is of course a physical nonsense. Taking $l=3.5 a=1.75 D$, as in the parameter set $\mathcal{R}'$, avoids any singularity in all the cases studied in the present work. At short, nanometric distances, a more realistic theory is in principle required, which better takes into account the discrete nature of lipids. Their ability to tilt in order to relax a boundary constraint might for example help solving this issue~\cite{Fournier98,Bohinc03}. But such a refinement is out of the scope of the present work. 

We thus adopt the following approach: we compare (in Appendix~\ref{Nbody}) three-body truncations and exact calculations for $l=3.5$ and we demonstrate that three-body truncations are good approximations. Consequently, as discussed in the Appendix, when $l=3.5$ we use exact results, whereas we limit ourselves to three-body truncations when  $l<3.5$.

\subsection{On the many-body character of Casimir and elastic interactions}
\label{manybody}

\begin{figure}[ht]
\begin{center}
\includegraphics*[width=7cm]{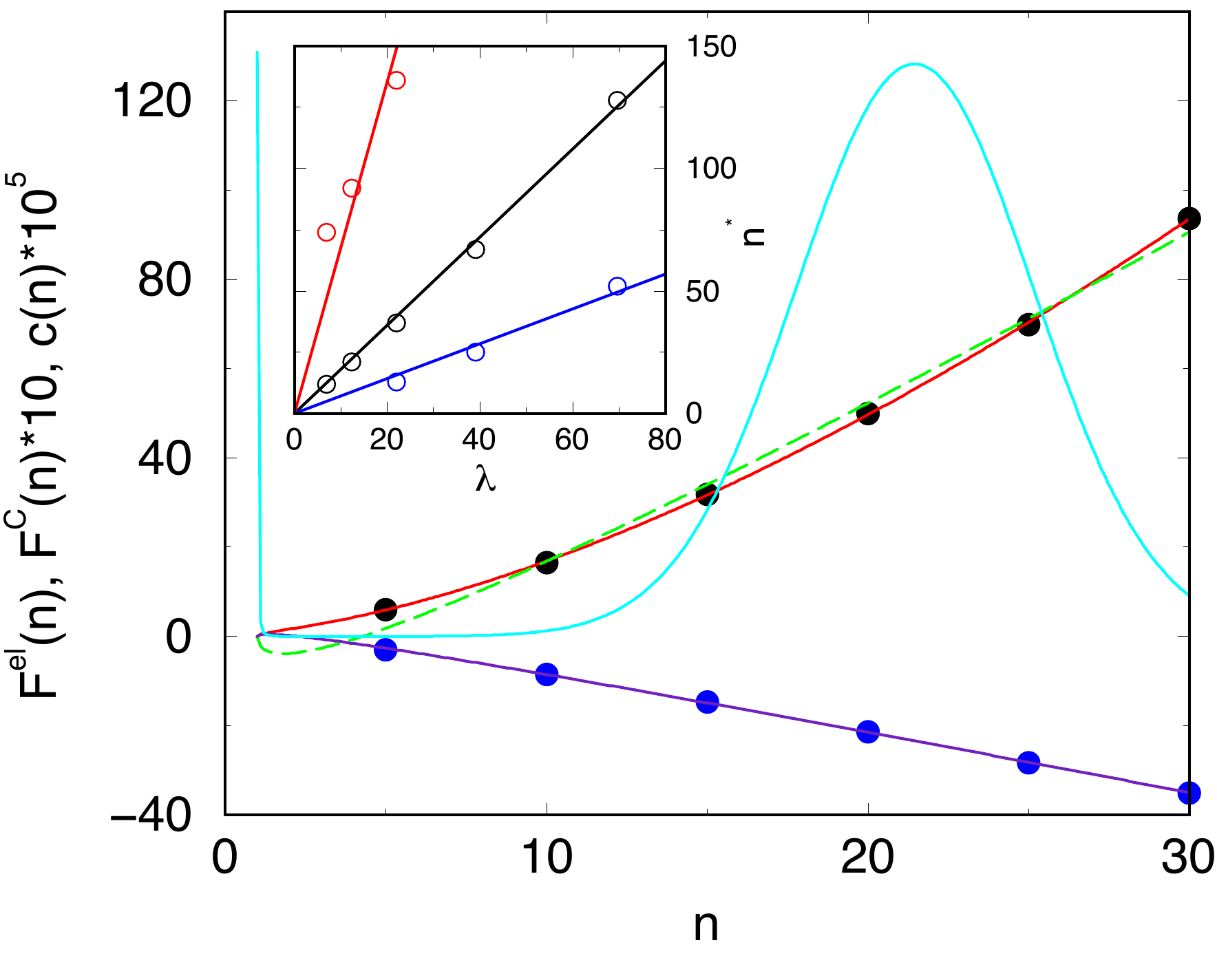}
\end{center}
\caption{Purely elastic energy $F^{\rm el}(n)$ (positive values, black dots) and Casimir energy $F^{\rm C}(n)$ (negative values, blue dots) for circular $n$-clusters. We used the $\mathcal{R}'$ parameter set (the membrane tension is $\sigma=1$ thus the membrane-propagated interaction range is $\lambda=12.4$) and we performed $R=1000$ realizations in each case (error bars are smaller than symbol size). Lines are the best fits on the interval $n\in[5,30]$. Dashed lines correspond to the simple droplet model whereas continuous ones correspond to the generalized model, Equation~(\ref{droplet:form}); dashed and continuous lines are superimposed for the Casimir energy. There is no positional randomness in the clusters (see text). In cyan, the resulting cluster-size distribution $c(n)$ for $f_0=\rho_0=12 k_{\rm B}T$ and an inclusion surface fraction $\phi=0.25$. Inset: Values of the most probable cluster size $n^*$ in function of $\lambda$ (in units of $a$) for the same parameters except that, from bottom to top, inclusion contact angles are $\gamma=0.1$, $0.2$ and $0.4$~rad. These values of $\lambda$ correspond to $0.032 \leq \sigma \leq 3.2$ (i.e. $3.2 \times 10^{-5} \leq  \sigma  \leq 3.2 \times 10^{-3}$~J.m$^{-2}$  in real units). Lines are the best linear fit, from bottom to top $n^*=6.78 \lambda$, $n^*=1.80 \lambda$ and $n^*=0.71 \lambda$.  $R\geq 200$ in all cases.
\label{Fden}}
\end{figure}

The expressions of the block sub-matrices of $M$, resulting from equations~(\ref{propag}) and (\ref{diag}), reveal that the Casimir term $F^{\rm C}$ is a (complex) combination of Bessel functions $K_\nu(r_{ij}/\lambda)$, decaying exponentially with $r$, and of large inverse powers of the same $r_{ij}/\lambda$. Considering a given bulk inclusion at position $\mathbf{r}_{i_0}$, and summing over all the remaining inclusions, $j\neq i_0$, in a large circular $n$-cluster, it is thus reasonable to expect the sum to have a finite value when $n$ goes to infinity. In other words, we expect $F^{\rm C}(n)$ to scale like $n$, that is to say to be extensive, with leading corrections of order $\sqrt{n}$ due to particles close to the boundary. It is indeed what we observe. Fitting $F^{\rm C}(n)$ with the expected simple droplet form $-f_1(n-1)+\rho_1 \sqrt{n-1}$~\cite{Foret08}, without the repulsion term, yields a very accurate fit, as exemplified in Figure~\ref{Fden} for the $\mathcal{R}'$ parameter set. In this $\mathcal{R}'$ case, when fitting on the interval $n\in[5,30]$, the effective Casimir term is $F^{\rm C}(n)\simeq - 0.15 (n-1) + 0.18 \sqrt{n-1}$. 

The absence of the repulsive, third term in this droplet form [compare with Eq.~(\ref{droplet:form})] means that the Casimir attraction alone cannot be responsible for any cluster phase. This could be anticipated, because it is known to be attractive at the two-body level, whereas a repulsive force is required to stabilize cluster phases. However, some repulsive term might have been hidden in the higher-order many-body terms. This is evidently not the case. Even though it does not lead to any repulsive term, this Casimir interaction renormalizes the ``bare'' parameters $f_0$ and $\rho_0$ that would result from the short-range attraction alone (see Section~\ref{MicelTheo}). We note $f_{\rm r}\equiv f_0+f_1$ and $\rho_{\rm r}\equiv \rho_0+\rho_1$ the renormalized parameters of the droplet theory. From the above fits and the absolute values of $F^{\rm C}$, it thus appears that the Casimir term $F^{\rm C}$ eventually plays a negligible role since it renormalizes (i) $f_0$, but $f_0$ is irrelevant in the grand-caconical ensemble where it is compensated by the chemical potential; (ii) $\rho_0$, but the bare value of $\rho_0$ is very large and modifying slightly it has little effect according to anterior analytical work~\cite{Foret08}.

As a first consequence, above their critical concentration $\phi^{\rm c}$, (up-down) {\em symmetric} inclusions, the interaction of which reduces to the Casimir term since $\gamma_k\equiv0$, will condense at equilibrium in a macroscopic cluster, coexisting with a gaz phase (monomers and rare small multimers).

As for the elastic term $F^{\rm el}$, even though one might expect the previous argument to remain valid, its behavior is in fact more complex. As exemplified again in 
Figure~\ref{Fden}, the previous simple droplet model does not fit perfectly the $F^{\rm el}(n)$ data. However, as motivated in Section~\ref{MicelTheo} above, introducing in $F^{\rm el}(n)$ the additional, non-extensive term $\chi (n-1)^\alpha$ with $\chi >0$ and $\alpha>1$, which reflects a long-range repulsion between inclusions, significantly improves the fit. For example, in the $\mathcal{R}'$ case, when fitting on the interval $n\in[5,30]$, the effective free energy is $F^{\rm el}(n)\simeq - 0.87 (n-1) + 1.80\sqrt{n-1} + 0.75 (n-1)^{1.48}$. Note that this implies not only the existence of a long-range repulsion term in the total free energy $F(n)$ but also a renormalization of the parameters $f_0$ and $\rho_0$ as in the Casimir case above. More precisely and quite surprisingly, the renormalization goes into the direction of an {\em increase} of $f_{\rm r}$ and $\rho_{\rm r}$ as compared to $f_0$ and $\rho_0$, and thus of an {\em increase} of the short-range attraction.

\begin{figure}[t]
\begin{center}
\includegraphics*[width=7cm]{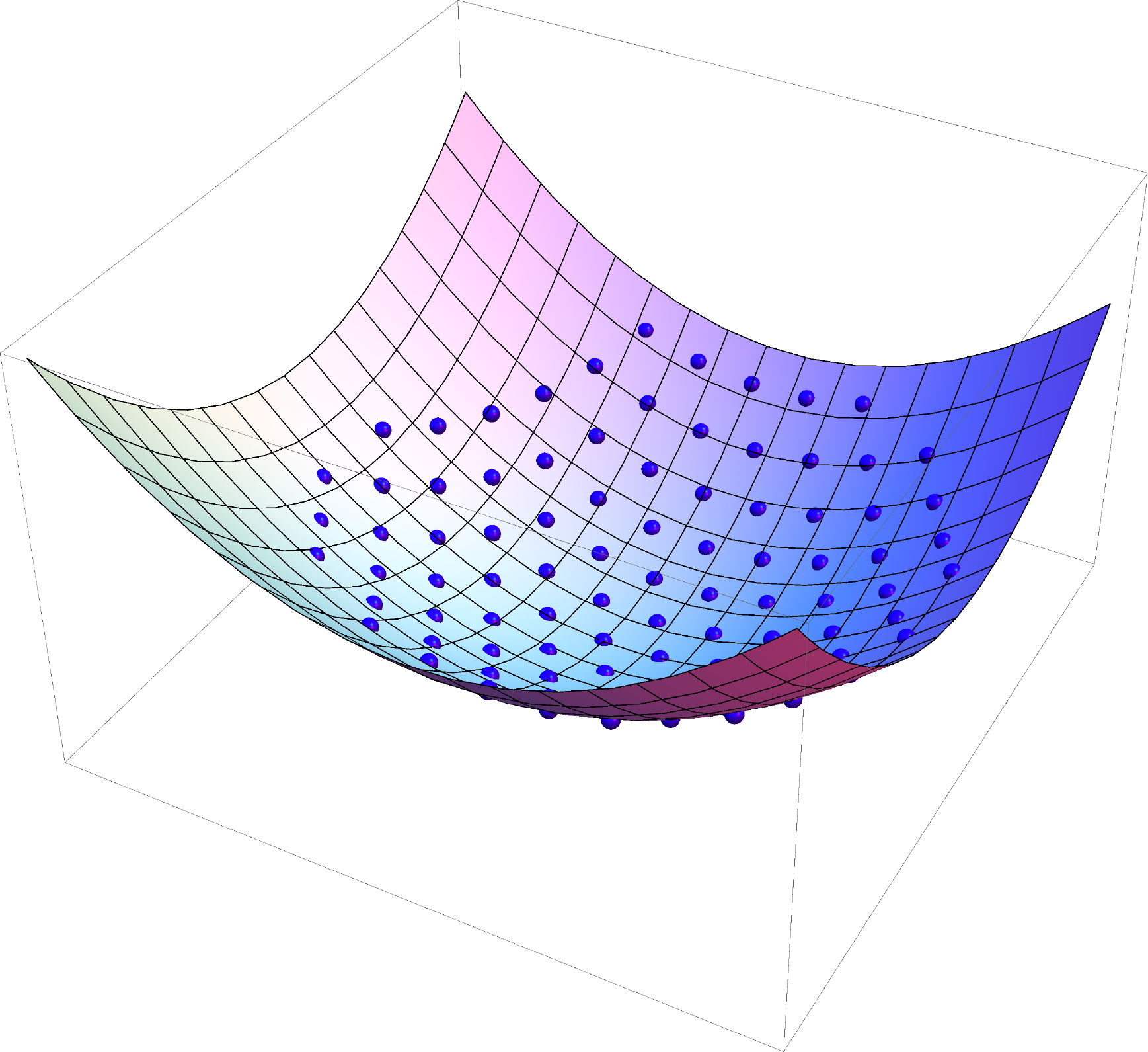}
\end{center}
\caption{Cluster of $n=91$ identical inclusions, $\mathcal{R}''(39.1)$ parameter set (i.e. $\sigma=0.1$). The blue dots represent the position of the inclusions calculated from Equation~(\ref{av:shape}). The represented surface is a quadratic fit, which shows that inside the cluster, the average membrane shape is very well approximated by a paraboloid of revolution of equation $z=\langle u(x,y) \rangle =\bar{C}(x^2+y^2)/2$ with $\bar C\simeq0.052$.
\label{Fit:profile}}
\end{figure}

The physical interpretation of the non-extensivity is as follows. At low tension $\sigma$, Helfrich's free energy~(\ref{eq:F-partielle}) is dominated by its curvature term and the tension term can be treated as a perturbation. Identical inclusions in the cluster impose a given mean curvature $\bar{C}$ inside the cluster (depending both on their contact angle $\gamma$ and the lattice spacing $l$). Minimizing Helfrich's free energy results, at order 0 in the perturbation, in a parabolic membrane invagination inside the cluster $\langle u(x,y) \rangle=\bar{C}(x^2+y^2)/2$ (if the cluster is centered at the origin, without loss of generality) \footnote{In the present small-gradient-of-$u$ approximation, a paraboloid of revolution is identical to a spherical cap, as expected for a regular assembly of inclusions imposing the same curvature.}. It is indeed what we observe at low tension, as illustrated in Figure~\ref{Fit:profile}. If $r_0\propto \sqrt{n}$ is the cluster radius, Helfrich's free energy inside the cluster then reads 
\begin{eqnarray}
F_{\rm cl.} & =& \int_0^{r_0} \left(\frac{\sigma}2  \bar{C}^2 r^2 + \frac{\kappa}2  \bar{C}^2 \right) 2\pi r {\rm d}r \\
 & =& \pi\bar{C}^2 \left(  \sigma \frac{r_0^4}4 + \kappa \frac{r_0^2}2  \right) \label{Fquad} 
\end{eqnarray}
The first term in the parentheses is subdominant since $\sigma$ is supposed to be small, but it is proportional to $n^2$ and thus makes the total membrane free energy~-- adding the contributions inside and outside the cluster~-- non-extensive, as required, with $\alpha=2$. 

In the high or intermediate tension regime, the situation is more involved, because minimizing the free energy will result in a membrane shape more or less deviating from a paraboloid inside the cluster. In practice, we indeed observe that the exponent $\alpha$ is different from 2, since it is close to 1.5 for $\sigma=1$. However, when $\sigma$ decreases, we also observe that $\alpha$ gets closer to 2, as expected. For example, for $\sigma=0.1$, $\alpha \simeq 1.7 \approx 2$. Finally, the values of the effective curvature in the cluster, $\bar C$ are as follows: for $\sigma=1$, 0.1 and 0.01, $\bar C=$0.035, 0.052 and 0.057, respectively. These values are discussed in Appendix~\ref{barC}.

The previous analysis relies on inclusions with fixed positions inside clusters, since they dwell on a regular, triangular lattice. Nevertheless, at finite temperature, positional fluctuations occur and should in principle be taken into account. However, we have checked that they play a subdominant role. Indeed, in first approximation, they can be modeled by adding a small random contribution $\mathbf{r}$ to each inclusion position inside a cluster, with $|\mathbf{r}| \ll l$. More precisely, the probability distribution of $\mathbf{r}$ has been chosen to be a Gaussian $p(\mathbf{r})=1/(2\pi s^2)\exp[-\mathbf{r}^2/(2 s^2)]$ of standard deviation $s=a/16$ or $a/8$. We have checked on representative parameter sets [$\mathcal{R}'$, $\mathcal{R}''(22.0)$ and $\mathcal{R}''(39.1)$, corresponding respectively to membrane tensions $\sigma=1$, $1/\sqrt{10}$ and 1/10] that this has marginal influence on the observed values either of $F^{\rm C}(n)$ and $F^{\rm el}(n)$, or of $n^*$, for both values of $s$. Thus we shall not consider this possibility any longer in the sequel of the paper.

To finish with, we now focus on the $\mathcal{R}$ parameter set where we have to limit ourselves to three-body truncations because $l=2$. We also find cluster phases in this case, with $n^*=21$ when $\phi=0.25$. As anticipated, this value is identical to the $\mathcal{R}'$ case one, because we have rescaled proportionally the lattice spacing $l$ and the repulsion range $\lambda$. When the tension is reduced to $\sigma=0.1$, $n^*=71$ is also quite close to the $\mathcal{R}''(39.1)$ (i.e. $\sigma=0.1$) case value, $n^*=69$. This reinforces the conclusions of Appendix~\ref{Nbody} that three-body truncations reproduce correctly the whole many-body repulsive interaction, at least as far as identical inclusions are concerned.

To conclude this section, we have enlightened the collective character of the elastic interaction, related to the intrinsically collective nature of the budding process, i.e. of the large-scale deformation of the membrane (Figure~\ref{profile1}). By contrast, the Casimir interaction is the same as in the symmetric case ($\gamma_k \equiv 0$) where the membrane remains flat on average~-- up to thermal fluctuations~-- and remains extensive. The non-extensivity of $F^{\rm el}$ will be of great interest in the following. 

\subsection{Cluster phases}

Having clarified the many-body character of the long-range repulsion, we can now focus on the cluster phases under interest, which result from a balance between this repulsion and the short-range attraction. To begin with, we still consider in this section an homogeneous population of inclusions with the same contact angle, $\gamma_k\equiv \gamma={\rm Const.}$, before investigating the coexistence of inclusions imposing opposite curvatures in Section~\ref{opposite}. We adopt the statistical mechanics approach of Reference~\cite{Foret08}, as presented in Section~\ref{theo}: we calculate the cluster-size distribution $c(n)$ by using the fitted functions $F^{\rm el,C}(n)$ of the cluster free energies that we have just obtained above and applying the approximate analytical scheme developed in Section~\ref{MicelTheo}. Note that it might happen that the fit parameters $f_1$, $\rho_1$, $\chi$ or $\alpha$ depend on the choice of the fitting interval $[n_{\rm min},n_{\rm max}]$, notably on $n_{\rm max}$ ($n_{\rm min}$ is always chosen close to 1). However, we have systematically verified that the ensuing most probable cluster-size, $n^*$, does not depend on $n_{\rm max}$ provided that $n^*<n_{\rm max}$.

Figure~\ref{Fden} illustrates the procedure by showing, for the $\mathcal{R}'$ parameter set, the numerical values of $F^{\rm el}$ and $F^{\rm C}$ in function of $n$, the fits on the interval [5,30], and the cluster-size distribution (beware of the scales on the $y$-axis). Cluster phases are stable at equilibrium because the distribution is bimodal: monomers co-exist with multimers of typical (or most probable) size $n^*$. This is coherent with anterior works that did not take many-body potentials into account~\cite{Sear99,Bibi08,Foret08,Meilhac11}. We demonstrate in Appendix~\ref{Nbody} that this is due to the fact that $N$-body forces for $N>3$ only contribute marginally to the cluster free energy.

The Inset of Figure~\ref{Fden} also shows this most probable cluster size $n^*$ in function of $\lambda$ in the $\mathcal{R}''(\lambda)$ parameter set (black symbols, $\gamma=0.2$~rad) and for two additional values of the inclusion contact angles $\gamma=0.1$, 0.4~rad~\footnote{When varying parameters, one must systematically keep sure that $n_{\rm max}$ remains larger than $n^*$.}. It clearly demonstrates that $n^* \propto \lambda \propto 1/\sqrt{\sigma})$ for the studied values of $\lambda$ (two decades) and $\gamma$. This observation will have to be given a theoretical explanation in future investigations. Furthermore, at the studied fraction $\phi=0.25$, the respective heights of the monomer and multimer peaks remain of the same order of magnitude, which proves that one keeps being in a true cluster phase even at large $\lambda$. Together, these results suggest that even though the typical cluster-size increases with decreasing surface tension $\sigma$, a cluster phase survives even at very low surface tension. This is of particular interest in the biological context of plasma membrane proteins, because cell membrane tensions are well below the lysis tension~\cite{Evans76,Sheetz96}. Finally, the behavior of $n^*$ in function of $\gamma$ confirms that the weaker the repulsion, the larger the clusters~\cite{Foret08}.

Anterior studies~\cite{Bibi08} supposed that $|F^{\rm el}|$ had to be larger than $|F^{\rm C}|$ because in the vanishing tension limit, both quantities have the same $1/r^4$ asymptotic behavior at large separations $r$. If this inequality were not satisfied, then the attractive purely elastic attraction would dominate, and there would be no repulsion any more. This condition implied that $\gamma$ had to be large enough, typically $\gamma\gtrsim5^\circ$. The present analysis shows that this argument is erroneous in the finite tension case, of biological interest. Both $F^{\rm el}$ and $F^{\rm C}$ renormalize $f_0$ and $\rho_0$, but only $F^{\rm el}$ contributes (positively) to the repulsion parameter $\chi$, whatever the value of $\gamma$. Of course, the typical cluster size $n^*$ grows as $\gamma$ or $\sigma$ decrease, but the present study suggests that cluster phases always exist provided that $\gamma>0$ and $\sigma>0$. 

\subsection{About the zero tension case}
\label{sigma:0}

The observation that the cluster size diverges (proportionally to $\lambda$) when $\lambda \rightarrow \infty$, i.e. when $\sigma \rightarrow 0$, suggests that when $\sigma=0$, above the critical surface fraction $\phi^{\rm c}$~\cite{Foret08}, the condensed phase consists of a large, macroscopic cluster. In this zero tension case, when calculating $F^{\rm el}(n)$ as in the previous section, one consistently observes that fitting $F^{\rm el}(n)$ with a modified droplet form with a repulsion term [Equation~(\ref{droplet:form})] does not significantly improve the fit quality as compared to the simple droplet form without this term (data not shown). This absence of repulsion term also means that the condensed phase is a macroscopic cluster above $\phi^{\rm c}$. 

This implies in principle that no cluster phase exists in this limit, which is in contradiction with the conclusions of Reference~\cite{Meilhac11}. However, the present analysis relies on the underlying assumption that clusters are compact objects with a disc-like geometry while the clusters observed in this anterior numerical work presented a more elongated, possibly fractal shape when $\sigma=0$. Therefore the present analysis might not be adapted to this situation and will deserve a separate treatment in a future work, possibly because of the divergence of the interaction range $\lambda$, which might invalidate the no-interacting cluster approximation in section~\ref{MicelTheo}. Note also that in contrast to the present case, the cluster-size distributions observed in this reference at vanishing tension were not bimodal but had a power-law tail as in percolation clusters, also indicating that a different mechanism should be at work.

\section{Coexistence between inclusions of opposite orientations}
\label{opposite}

\begin{figure}[!t]
\begin{center}
\includegraphics*[width=7cm]{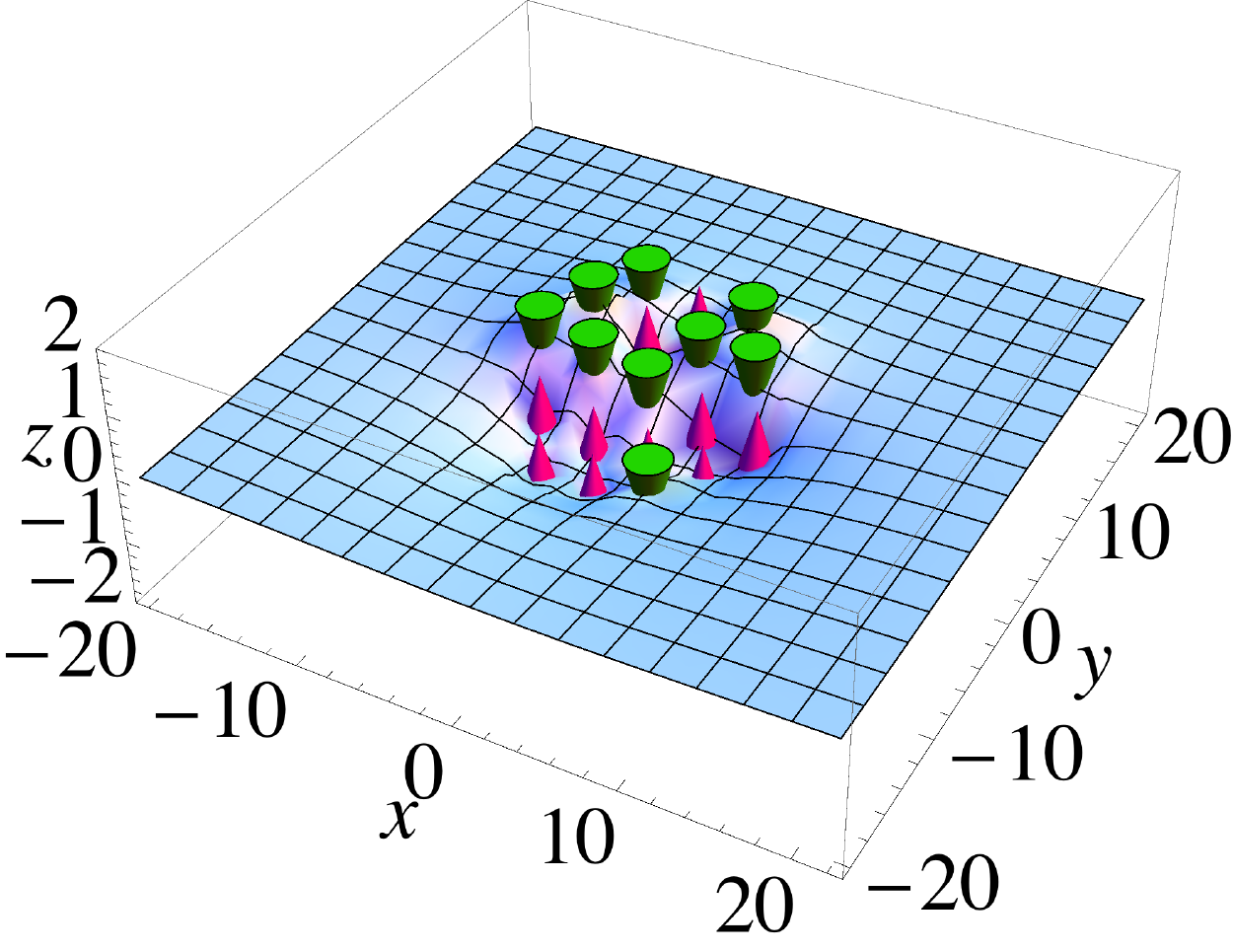}
\end{center}
\caption{Cluster of $n=19$ inclusions of opposite orientations arranged in a random fashion ($x_\uparrow\simeq0.5$) and consecutive membrane shape $z=\langle u(x,y) \rangle$ as calculated from Equation~(\ref{av:shape}). Inclusions imposing a positive (resp. negative) curvature are represented by pink (resp. green) cones. We use the $\mathcal{R}'$ parameter set but the overall aspect would be the same at lower tension:  as compared to Figure~\ref{profile1} where the cluster only contained inclusion having the same orientation, the membrane does not present a globally budded aspect anymore. 
\label{profiles2}}
\end{figure}

As exemplified in Figure~\ref{profiles2}, the coexistence of inclusions of opposite orientations in a same cluster makes the membrane shape more irregular at short distances, whereas it does not present any global invagination.
We thus expect clusters mixing inclusions of opposite orientations to behave differently from their homogeneous, same-orientation counterpart as discussed so far. We indeed demonstrate below that in the moderate membrane tension regime of biological interest ($\sigma \approx 0.1$, i.e. $10^{-4}$~J/m$^2$), the free energy of such a mixed cluster is less favorable than its homogeneous counterpart. A a consequence, inclusions of opposite orientation phase-separate in distinct clusters. This is an original curvature-driven demixing mechanism. At higher membrane tensions, the conclusion will depend on the lattice spacing $l$.

We assume that only two inclusion species coexist, with equal absolute contact angles $\gamma$. These inclusions are still assumed to be isotropic. We denote by $x_\uparrow$ (resp. $x_\downarrow=1-x_\uparrow$) the fraction of inclusions of contact angle $+\gamma>0$ (resp. $-\gamma<0$) in a given cluster; $x_\uparrow$ and $x_\downarrow$ play a symmetrical role. We again focus on $\gamma=0.2$~rad.

\begin{figure*}[ht]
\begin{center}
\includegraphics*[width=7cm]{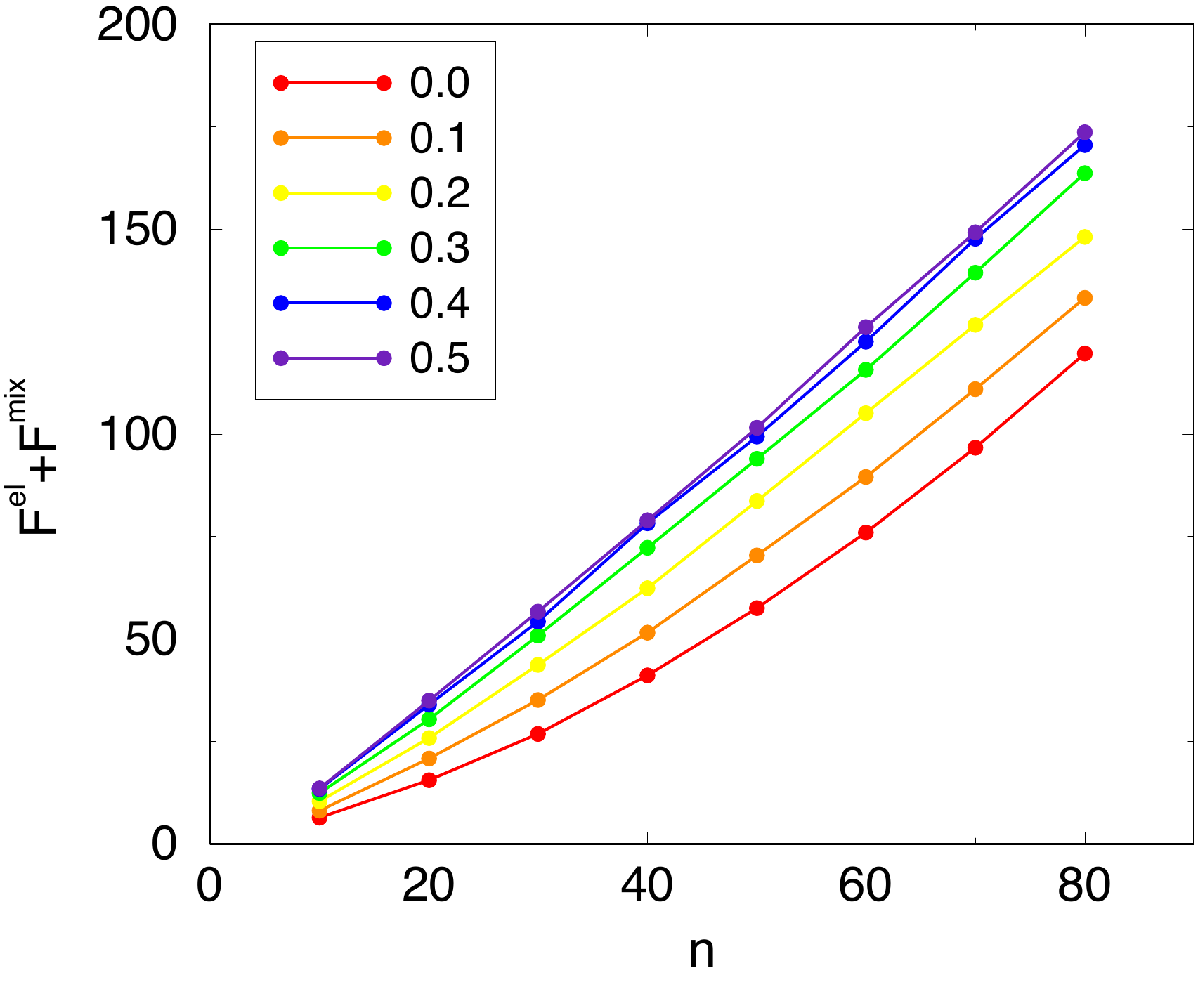} \qquad
\includegraphics*[width=7cm]{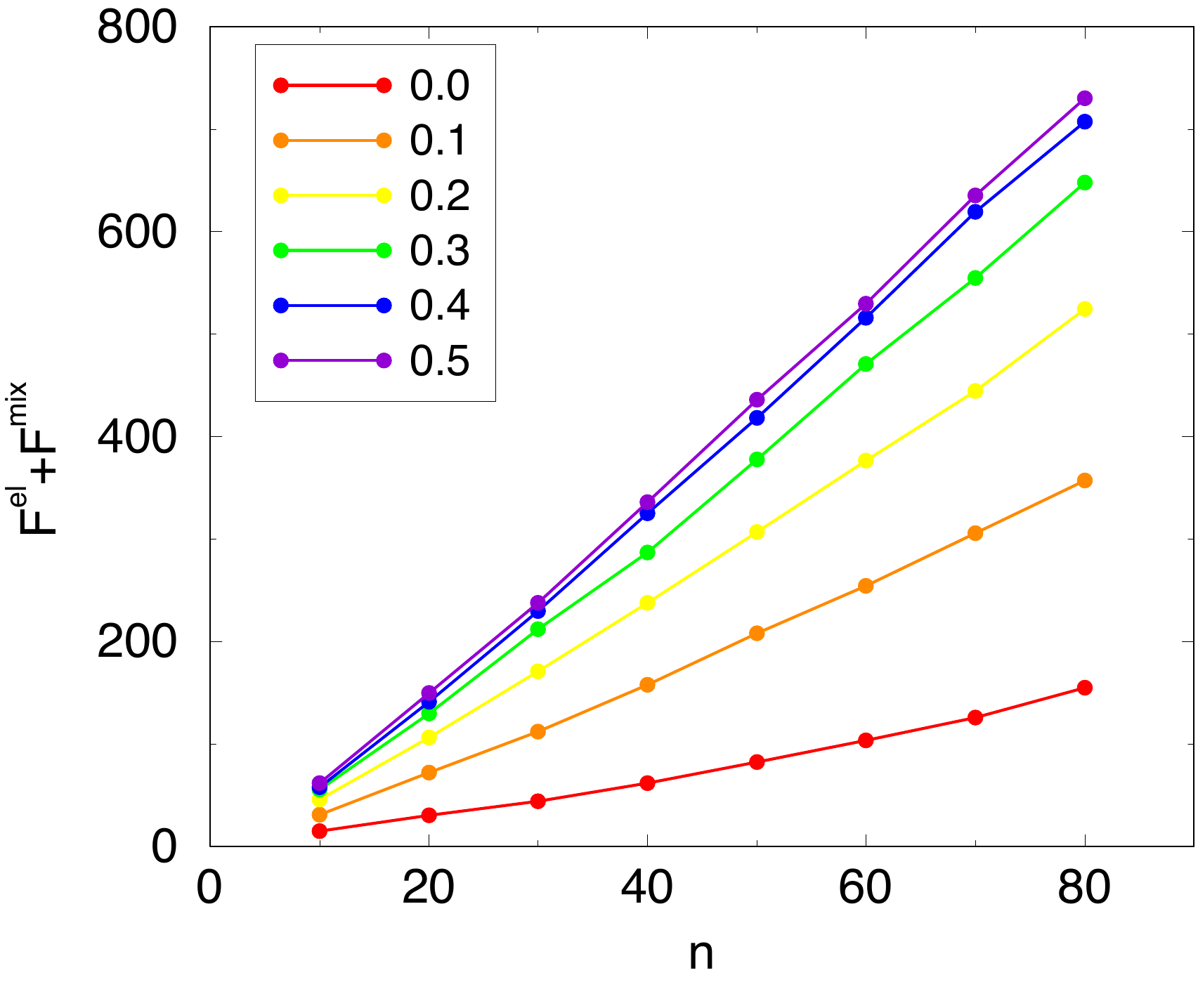}
\end{center}
\caption{Free energies $F^{\rm el}(n)+F^{\rm mix}(n,x_\uparrow)$ of clusters of size $n$ containing inclusions of opposite orientations, the fraction $x_\uparrow$ (or, symmetrically, $x_\downarrow$) being given in the figure legend. Left: $\mathcal{R}''(39.1)$ parameter set (i.e. $l=3.5$ and $\sigma=0.1$), exact calculation, $R=200$; Right: $\mathcal{R}$ parameter set (i.e. $l=2$) except that $\sigma=0.1$, three-body approximation, $R=100$. Error bars are smaller than symbol size. 
\label{rel:stability}}
\end{figure*}

Figure~\ref{rel:stability} displays the elastic free energies of clusters of a given size $n$, that contain inclusions with both orientations, with given fractions $x_\downarrow$ (or, symmetrically, $x_\uparrow$), randomly distributed inside the circular cluster. The tension is moderate, $\sigma=0.1$. To compare the relative stabilities of both phases, one must in principle also take into account (i) the bare short-range contribution $F^{\rm s.r.}$ (i.e. the ``bare'' parameters $f_0$ and $\rho_0$): to start with, we assume them to be equal in the homogeneous and mixed cases; (ii) the Casimir entropies: they do not depend on the angles $\gamma$ and they are thus identical in both cases; (iii) and the mixing free energy~\cite{Bibi10} 
\begin{equation}
F^{\rm mix}(n,x_\uparrow) = n k_{\rm B}T (x_\uparrow \ln x_\uparrow + x_\downarrow \ln x_\downarrow).
\end{equation}
Following the same procedure as in Reference~\cite{Bibi10}, one then has first to minimize the total free energy $F(n,x_\uparrow)=F^{\rm el}+F^{\rm C}+F^{\rm mix}+F^{\rm s.r.}$ with respect to $x_\uparrow$. In both cases displayed in Figure~\ref{rel:stability}, this minimization with respect to $x_\uparrow$ shows that the most favorable case is the demixed one independently of $n$, i.e. $x_\uparrow=0$ or $x_\uparrow=1$. One then applies the same procedure as previously, 
which eventually leads to the same most probable cluster size $n^*$ as in the homogeneous case.

Consequently, it appears that in the moderate tension regime of biological interest, due to membrane-propagated interactions alone, and independently of the short-range contribution, proteins of opposite orientation phase-separate~\cite{Bibi10} in distinct clusters. The consequences in terms of membrane budding will be discussed in the conclusion.

The behavior in the high-tension regime is more complex because the conclusion then depends on the inter-inclusion distance $l$. At $l=2$, the behavior is comparable to the moderate-tension one, whereas at $l=3.5$ phase-separation does not occur (see appendix~\ref{demix:high} for additional information and a mechanism accounting for this observation). 

Furthermore, it appears that when inclusions of opposite orientations are equally mixed ($l=2$ or $l=3.5$), $F^{\rm el}$ is again very well fitted by a simple droplet model, without repulsion term, as $F^{\rm C}$ above. We attribute this observation to the fact, illustrated in Figure~\ref{profiles2}, that the membrane is not invaginated anymore because of the long-range ``screening" due to the inclusions imposing opposite curvatures. On average, the membrane is flat a long distances, as in the symmetric inclusion case in Section~\ref{manybody}, and one recovers extensivity of the purely elastic interaction. Again, extensivity implies that cluster phases are not stable and that mixed inclusions condense in a large, macroscopic cluster. In the large tension and large lattice spacing case, we conclude that if one wants to prevent condensation of (close to) equally-mixed inclusions in a macro-phase, there must exist short-range attractive forces, now depending on the nature of the interacting inclusions, which favor the aggregation of inclusions of same orientation, as in References~\cite{Bibi10,Meilhac11}.

\section{Conclusion and discussion}
\label{cl}

In previous studies based on mesoscopic-scale Monte Carlo simulations, the collective organization of inclusions in an elastic membrane has been studied using ad-hoc expressions of the membrane-mediated free energy~\cite{Bibi08,Meilhac11}. A goal of the present contribution was to investigate this question more rigoruously, using free-energy expressions derived from a realistic physical model of the membrane. Even though both approaches generally lead to the same qualitative conclusion (namely the existence of cluster phases), the present approach shed new light on the mechanisms at work. In some cases, they can also lead to different conclusions:  interestingly, the obtained results differ for the zero-tension case (see Section~\ref{sigma:0}). The Monte Carlo study concluded to the existence of a cluster phase, whereas our calculations lead to a single macroscopic cluster (no cluster phase). However we cannot conclude definitively because our calculations rely on other assumptions (e.g. that clusters are circular). This question may be answered in future studies by performing Monte Carlo simulations using, at each simulation step, the free-energy expressions obtained here.

As stated in the Introduction, coarse-grained Molecular Dynamics (MD) simulations, where a particle represents a group of atoms, can also bring very useful insight into membrane inclusions organization, even though accessible time scales and system sizes make it difficult to tackle equilibrium statistical mechanics. To our knowledge, the only MD works studying assemblies of asymmetric inclusions presumably close to equilibrium are References~\cite{Reynwar07,Periole08,Reynwar08}. Their conclusions regarding membrane-mediated aggregation or curvature-induced budding are qualitatively consistent with ours. However, they do not provide sufficient statistical sampling to allow a quantitative comparison. The comparison to experimental results is even more delicate because, even though experimental techniques have dramatically progressed in the recent decades~\cite{NSOM,Lenne09,Lang10,Bogaart11,Schreiber12}, and demonstrate the existence of small protein clusters in cell membranes, it remains difficult to conclude that the present scenario is able to account for experimental results. Specific biophysical experiments will have to be designed in the future to tackle this issue.

In this paper, we have also put forward an original curvature-induced process for segregation of inclusions in a biomembrane: at moderate surface tension of biological interest, inclusions of opposite orientations are segregated in distinct clusters because the free energy corresponding to mixed clusters is bigger than that of pure clusters. However, we have also shown that at higher surface tension, the conclusion depends on the inter-inclusion distance, which merits further investigations. This original curvature-induced demixing process might by of particular interest from a biophysical perspective: when segregating in distinct clusters, protein of opposite orientation will promote buds or invaginations~\cite{Reynwar07}, pointing inwards or outwards, whereas they would not promote any invagination at all if they where mixed in clusters. Another extension of our work might be to consider systems mixing not only opposite orientations, but also inclusions imposing several different curvatures, including a fraction of up-down symmetric ones, and to investigate under which conditions phase-separation still occurs. This would be a more realistic setup for real biomembranes.

Our study of this demixing process has also highlighted the importance of many-body forces for the collective behaviour of the inclusions: as stressed in Appendix~\ref{Nbody}, only considering pairwise additive forces~\cite{Bibi08,Meilhac11}, even though leading to the correct qualitative conclusion, gives significantly smaller clusters than an exact calculation. Furthermore, when considering systems mixing inclusions of opposite orientations, previous calculations, considering only the  two-body truncation of the interactions~\cite{Weikl98}, concluded that no demixing occurs (because the two-body free energy corresponding to mixed clusters was lower than for pure clusters). Here we have shown that the conclusion radically changes when fully taking into account many-body interactions. Thus it seems that the demixing process relies essentially on many-body interactions. 

In this work, we used a continuous theory to describe the membrane in which the inclusions are considered as point-like particles, and this led us to free energy expressions that diverge for small inter-inclusion distances when considering the whole $N$-body interactions. Therefore we had to assume relatively high inter-inclusion distances in the clusters to be able to perform the full $N$-body calculations. We have checked that most of the conclusions are quite robust and do not depend on the inter-inclusion distance, by using the three-body truncations of the free energies (which never diverge). In addition, our analytical argument around Equation~(\ref{Fquad}) supports our findings independently of lattice spacings or divergence issues. However, some conclusions (in particular with regard to the demixing process at high surface tension, even though not of biological interest) depend strongly on the inter-inclusion distance. They should therefore be considered more carefully, as they are only supported be three-body calculations and an improved $N$-body calculation might alter our conclusions. It might also be interesting in the future to check these conclusions using a different approach in which the inclusions are not considered as point-like but as finite-size objects imposing conditions at their boundaries~\cite{Goulian93,Fournier97,Weikl98}.

Finally, a natural continuation of the present work will be to couple the inclusions to an additional field coding the lipidic concentration in the case of lipidic mixtures~\cite{Dean06}, in order to explore how proteins and lipids contribute in concert to the stability of membrane nano-domains having both a specific protein and lipid composition. This work is in progress.

\bigskip

\noindent{\bf Acknowledgements} 

One of us (ND) thanks the French Agence Nationale de la Recherche for temporary financial support (project ANR-09-PIRI-0008) and the University of Toulouse III and the French Centre National de la Recherche Scientifique for permanent support.

\appendix

\vspace{10cm}

\phantom{XXXXXXX}

\vspace{10cm}

\phantom{XXXXXXX}

\newpage

\centerline{\bf \large APPENDICES}

\bigskip

\section{Expansion of entropic Casimir interactions}
\label{purcasimir}

As far as Casimir interactions $F^{\rm C}$ are concerned, they do not depend on the $C_k$ nor on the $J_k$ [see Equation~(\ref{eq:F})] and are thus identical for symmetric, asymmetric and even anisotropic inclusions. The expansion in powers of $a/\lambda$ reads:
\begin{equation}
F^{\rm C}=F^{\rm C}_0+F^{\rm C}_{42}+F^{\rm C}_{62}+F^{\rm C}_{63}+\mathcal{O}\left[(a/\lambda)^8 \right],
\label{eq:FC-DLa6}
\end{equation}
where $F^{\rm C}_0$ is again a self-energy independent of the distances $r_{ij}$ and
\begin{widetext}

\begin{eqnarray}
F^{\rm C}_{42}&=&
\frac{a^4}{\lambda^4} \sum_{j>i} \Bigg\{
-576\frac{\lambda^8}{r_{ij}^8}
+576\frac{\lambda^6}{r_{ij}^6}{K_2^{(ij)}}
+24\frac{\lambda^4}{r_{(ij)}^4}\left[-3{K_0^{(ij)}}+4{K_2^{(ij)}}-6{K_2^{(ij)}}^2\right]
\nonumber\\ &&
\hspace{1.2cm}
+12\frac{\lambda^2}{r_{ij}^2}\left[3{K_0^{(ij)}}-4{K_2^{(ij)}}\right]{K_2^{(ij)}}
+\left[6{K_0^{(ij)}}{K_2^{(ij)}}-\frac{21}{8}{K_0^{(ij)}}^2-\frac{9}{2}{K_2^{(ij)}}^2\right] 
\Bigg\},
\end{eqnarray}
\begin{eqnarray}
F^{\rm C}_{62}&=&
\frac{a^6}{\lambda^6}\ln\frac{2\lambda}{a}\sum_{j>i}\Bigg\{
-576\frac{\lambda^8}{r_{ij}^8}+576\frac{\lambda^6}{r_{ij}^6}{K_2^{(ij)}}
+24\frac{\lambda^4}{r_{ij}^4}\left[-3{K_0^{(ij)}}+4{K_2^{(ij)}}-6{K_2^{(ij)}}^2\right]
\nonumber\\&&
+12\frac{\lambda^2}{r_{ij}^2}\left[3{K_0^{(ij)}}-4K_2^{(ij)}\right]K_2^{(ij)}
+\left[6{K_0^{(ij)}}{K_2^{(ij)}}-\frac{21}{8}{K_0^{(ij)}}^2-\frac{9}{2}{K_2^{(ij)}}^2\right]
\Bigg\}
\end{eqnarray}
and finally
\begin{eqnarray}
F^{\rm C}_{63}&=&
\frac{a^6}{\lambda^6}\Bigg\{
-\sum_{i>j>k}
\frac{3}{8}{K_0^{(ij)}}{K_0^{(ik)}}{K_0^{(jk)}}
\nonumber\\ &&
\hspace{0.9cm}
+\sum_{i>j,k}'
\Bigg[
-576\frac{\lambda^8}{r_{ik}^4r_{jk}^4}\cos(4\alpha^k_{ij})K_0^{(ij)}
\nonumber\\ &&
\hspace{2.3cm}
+12\frac{\lambda^4}{r_{ij}^4}\cos(2\alpha^i_{jk}-2\alpha^i_{ki}){K_2^{(ik)}}{K_2^{(jk)}}
-\frac{\lambda^4}{r_{ik}^2 r_{jk}^2}144\cos(4\alpha^k_{ij})K_0^{(ij)}K_2^{(ik)}K_2^{(jk)}
\nonumber\\ &&
\hspace{2.3cm}
+\frac{\lambda^2}{r_{ij}^2}{K_2^{ij}}\left[18\left(\cos(4\alpha^i_{jk})+\cos(4\alpha^i_{ki})\right){K_0^{(ik)}}{K_0^{(jk)}}-6\cos(2\alpha^i_{jk}-2\alpha^i_{ki}){K_2^{(ik)}}{K_2^{(jk)}}\right]
\nonumber\\ &&
\hspace{2.3cm}
+
3\left(\cos(4\alpha^i_{jk})+\cos(4\alpha^i_{ki})\right){K_0^{(ik)}}{K_0^{(jk)}}{K_2^{(ij)}}-\frac{9}{4}\cos(4\alpha^k_{ij}){K_0^{(ij)}}{K_0^{(ik)}}{K_0^{(jk)}}
\nonumber\\ &&
\hspace{2.3cm}
+\left(-\frac{1}{2}\cos(2\alpha^k_{ij})-4\cos(4\alpha^k_{ij})+\frac{3}{4}\cos(2\alpha^i_{jk}-2\alpha^i_{ki})\right)
{K_0^{(ij)}}{K_2^{(ik)}}{K_2^{(jk)}}
\nonumber\\ &&
\hspace{2.3cm}
-\cos(2\alpha^i_{jk}-2\alpha^i_{ki}){K_2^{(ij)}}{K_2^{(ik)}}{K_2^{(jk)}}\Bigg]
\nonumber\\ &&
\hspace{0.9cm}
+\sum_{i>j,k}'
12{K_0^{(ij)}}\cos(4\alpha^k_{ij})\left[
24\frac{\lambda^6}{r_{ik}^2 r_{jk}^4}{K_2^{(ik)}}+\frac{\lambda^4}{r_{jk}^4}\left(-3{K_0^{(ik)}}+4{K_2^{(ik)}}\right)-2\frac{\lambda^2}{r_{jk}^2}{K_2^{(ik)}}{K_2^{(jk)}}
\right]
\Bigg\}.
\label{FC43}
\end{eqnarray}

\end{widetext}

One can also check here that the vanishing-tension limit again coincides with the results of References~\cite{Domm99,Fournier03}. In particular, Casimir interactions are pairwise additive at the leading order $a^4/\lambda^4$~\cite{Domm99}, but not at higher orders.  

%
%
%

\section{Role of $N$-body forces, $N\geq3$}
\label{Nbody}

\begin{figure}[t!]
\begin{center}
\includegraphics*[width=7cm]{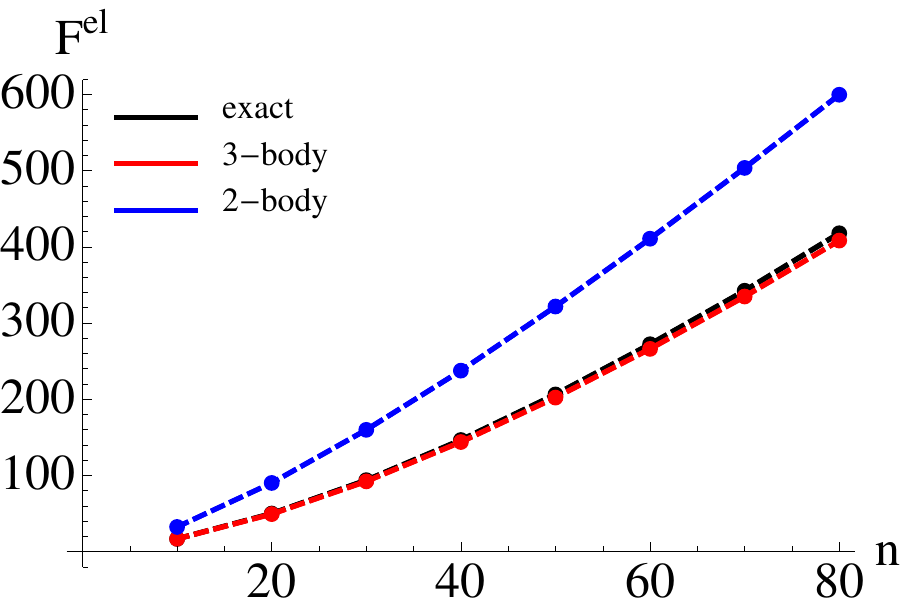}
\includegraphics*[width=7cm]{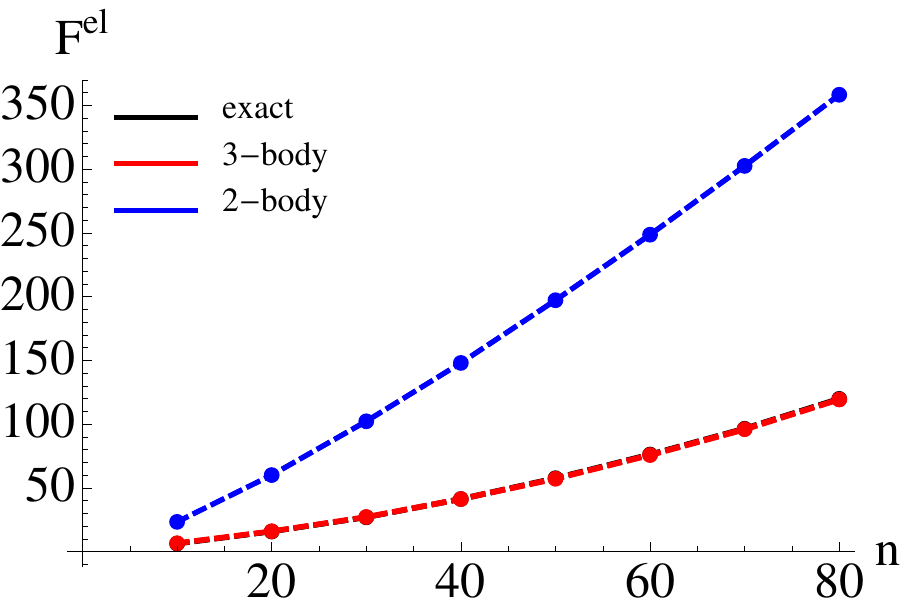}
\end{center}
\caption{Purely elastic free energy $F^{\rm el}$ in the case where all particles are identical, calculated with three different methods: exact calculations [Equation~(\ref{eq:F})], three-body truncation and two-body one. Top: $\mathcal{R}'$ parameter set. For $\phi=0.25$, the ensuing values of $n^*$ are 21 (exact and three-body) and 13 (two-body). Bottom: $\mathcal{R}''(39.1)$ (i.e. $\sigma=0.1$); the exact and three-body curves are superimposed. The ensuing values of $n^*$ are 69 (exact and three-body) and 38 (two-body) for $\phi=0.25$. $R=100$ in all cases.
\label{pur}}
\end{figure}

\begin{figure}[t!]
\begin{center}
\includegraphics*[width=7cm]{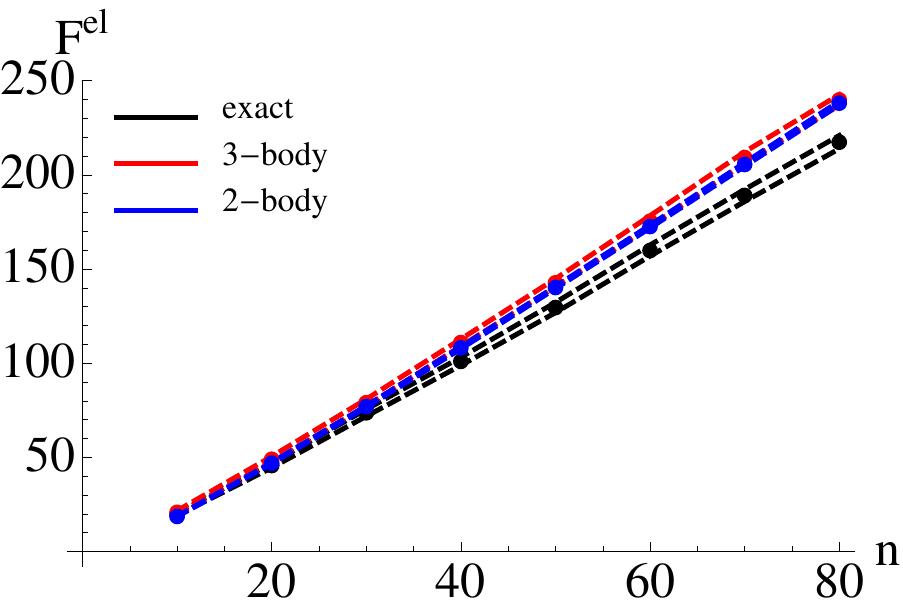}
\includegraphics*[width=7cm]{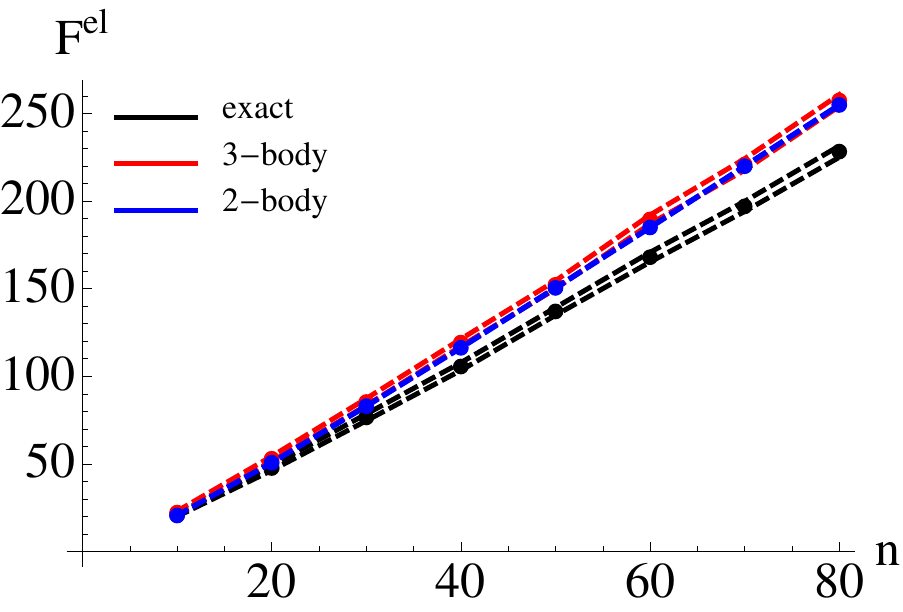}
\end{center}
\caption{Same as Figure~\ref{pur} except that clusters are a mixture of inclusions with opposite orientations, with $x_\uparrow=0.5$. Error bars are given to emphasize that in this case, up to error bars, three- and two-body approximations coincide.
\label{melanges}}
\end{figure}

Here we explore the role of $N$-body forces for clusters of inclusions with identical orientation (Figure~\ref{pur}), or where opposite orientations coexist (Figure~\ref{melanges}). To this end, when possible (i.e. when $l=3.5$ in order to avoid singularities), we compute the elastic free-energy $F^{\rm el}$ in three cases: exact calculations with the full matrix $M$, two-body, and three-body truncations. The results are displayed in the figures. They demonstrate that in the regimes of parameters explored in this work, (i) in the homogenous case (Figure~\ref{pur}), two-body truncations are manifestly insufficient to fully account for the forces mediated by the membrane, all the more so when $\sigma$ is low: the interaction is not pairwise additive. In contrast, a three-body truncation is sufficient:  it essentially provides the same free energy as the exact calculation and the ensuing values of $n^*$ are equal. (ii) In the mixed case (Figure~\ref{melanges}), three-body corrections do not improve significantly the two-body approximation; Both are a correct but not excellent approximation of the exact calculation. In any case, we assume that the three-body truncation is also a good approximation at shorter distances $l<3.5$.

This analysis stresses that numerical works only relying on pairwise interactions between inclusions~\cite{Sear99,Bibi08} are to be considered with care, even though correct at the quantitative level. Similarly, only focussing on two-body forces may lead to the wrong intuition that mixed clusters are systematically more stable than homogeneous ones~\cite{Weikl98}. A minimal requirement is to include three-body forces~\cite{Meilhac11}, but a totally correct approach consists of embracing the whole many-body character of elastic interactions mediated by the membrane, especially in the mixed case. We thus follow this approach when the lattice spacing is $l=3.5$. However, because of spurious divergences when inter-particle distances become too short ($l<3.5$), it is then useful to switch to the good three-body approximation. Conversely, the observed coincidence of exact calculations and three-body truncations observed at $l=3.5$ suggests that no such spurious effects should be induced, for this value of $l$, by the proximity of a divergence as discussed in Section~\ref{singul}.

\section{Values of the effective curvature $\bar C$ in clusters}
\label{barC}

In Section~\ref{manybody}, we have given the effective curvatures in clusters for different $\mathcal{R}''$ data sets: for $\sigma=1$, 0.1 and 0.01, $\bar C=$0.035, 0.052 and 0.057, respectively (in units of $a^{-1}$). The last value seems to be close to converged to the zero-tension limiting value discussed in the main text. In comparison, one can also calculate by a simple geometric argument the curvature $C'$ imposed by inclusions separated by a distance $l\geq a$, each of which imposes a contact angle $\gamma$: $C'=2\gamma/l$. For $l=3.5a$, we thus get $C'=2C/3.5=0.114a^{-1}$ since $C=0.2a^{-1}$ in the studied examples. Thus $C'=2\bar C$, whereas we expected both values to be equal. This shows that even though identifying $aC$ and the contact angle $\gamma$ is meaningful as far as long distance behaviors are concerned, as in References~\cite{Goulian93,Fournier97,Weikl98}, the two definitions do not match (by a factor two) when addressing short distances as in the present dense cluster. This stresses that a point-particle description is not {\em sensu stricto} equivalent to a finite-size particle one. In the present case, this amounts to considering particles whose effective contact angle at short distances is twice smaller than  $aC$. This does alter the overall significance of our conclusions. 

\section{Absence of demixing in the high tension and large lattice spacing regime}
\label{demix:high}

In this Appendix, we compare, in the high-tension regime ($\sigma=1$),  the relative stabilities of clusters mixing or not head-to-tail inclusions. We focus on the $\mathcal{R}$ and $\mathcal{R'}$ parameter sets. 
\begin{figure*}[ht]
\begin{center}
\includegraphics*[width=7cm]{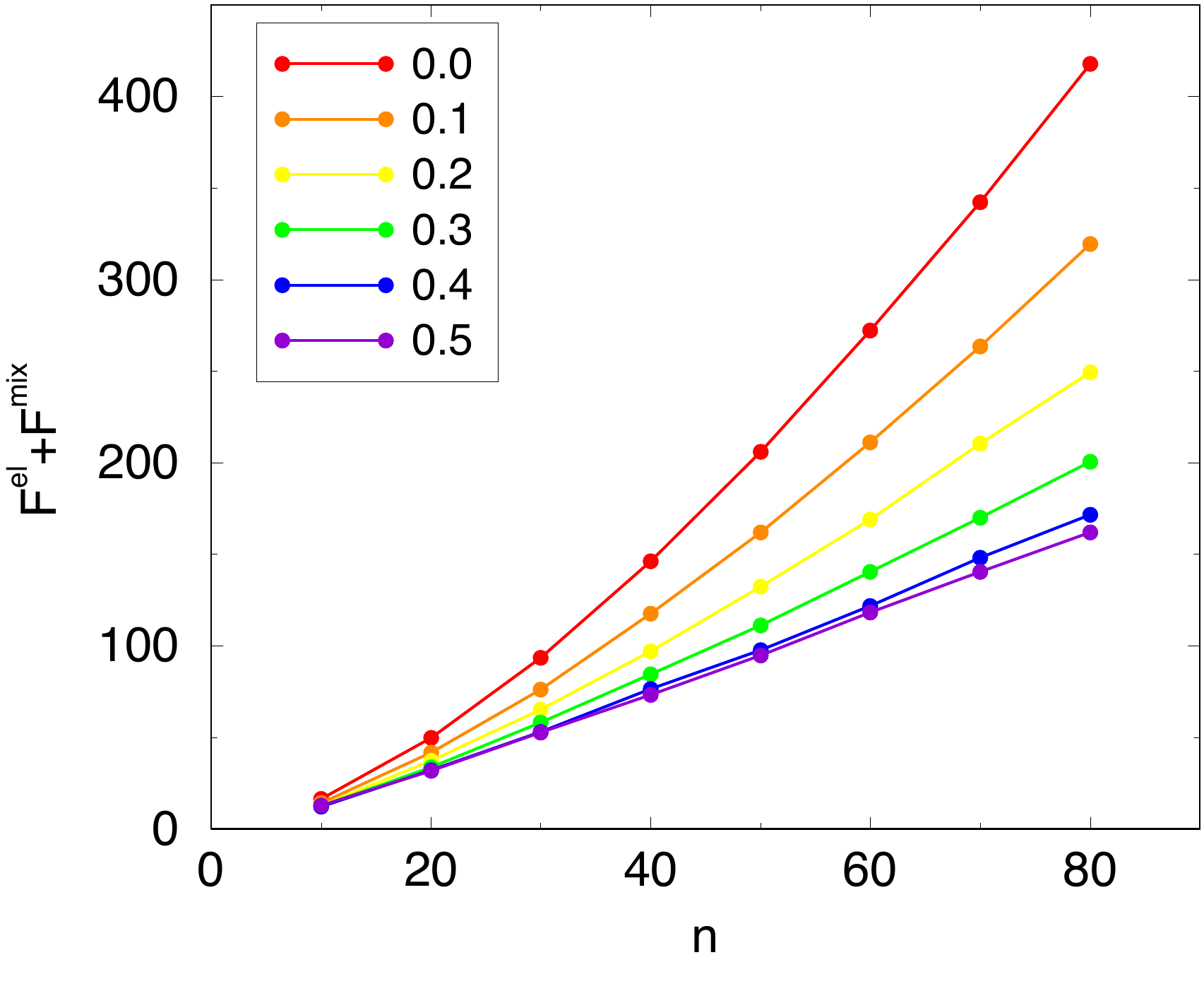} \qquad
\includegraphics*[width=7cm]{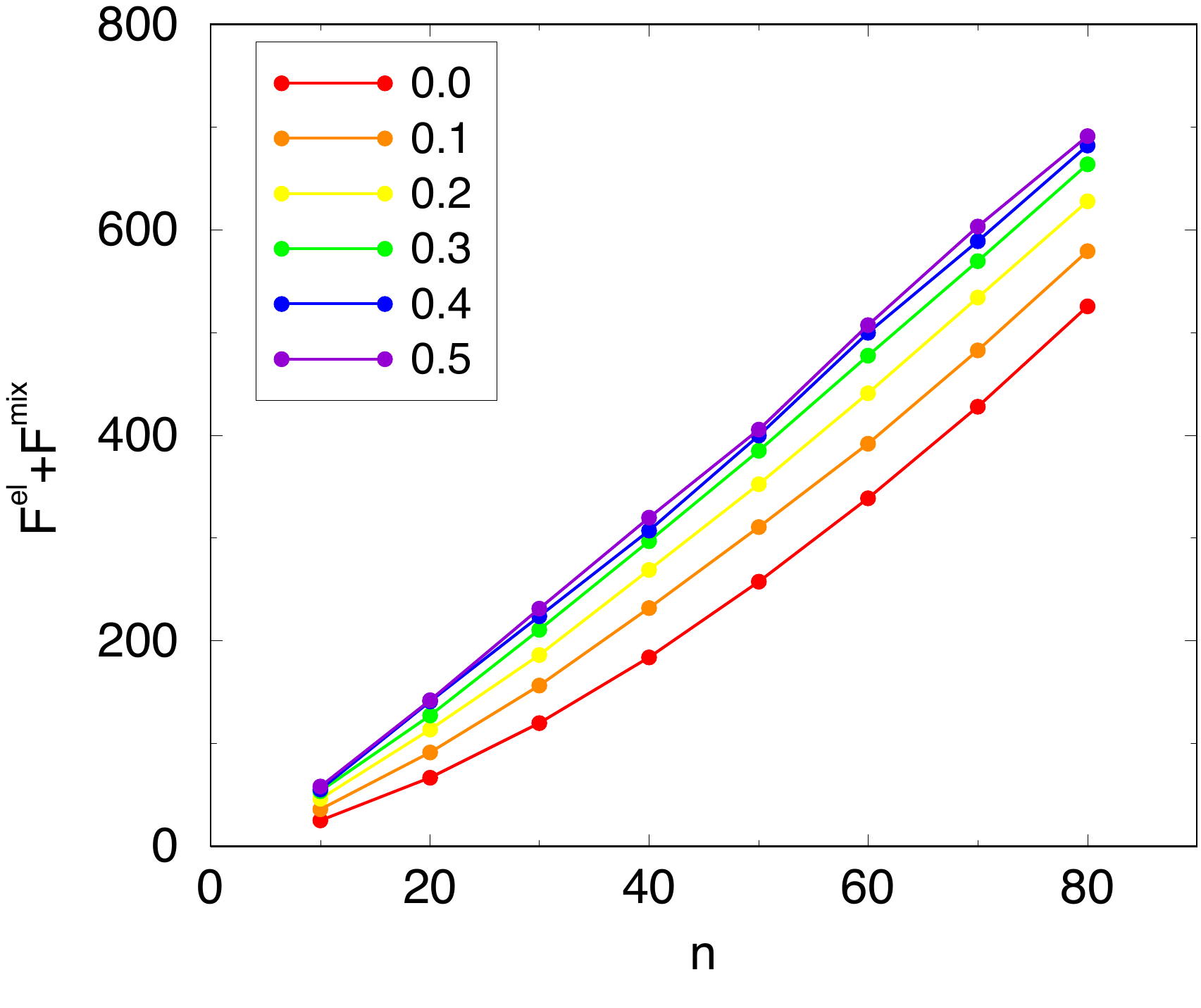}
\end{center}
\caption{Free energies $F^{\rm el}(n)+F^{\rm mix}(n,x_\uparrow)$ of clusters of size $n$ containing inclusions of opposite orientations, the fractions $x_\uparrow$ being given in the figure legend. Left: $\mathcal{R}'$ parameter set (i.e. $\sigma=1$ and $l=3.5$), exact calculation; Right: $\mathcal{R}$ parameter set (i.e. $\sigma=1$ and $l=2$), three-body approximation. Error bars are smaller than symbol size and $R=100$ in all cases. 
\label{rel:stability2}}
\end{figure*}

Figure~\ref{rel:stability2} shows that in this case, the relative stability depends on the lattice spacing $l$. At $l=2$, the behavior is comparable to the moderate-tension one ($\sigma=0.1$) discussed in the main text, whereas at $l=3.5$ phase-separation does not occur because the mixed case is more stable.

We interpret this finding as follows: when inclusions of opposite orientations are close, they impose antagonist conditions to the elastic membrane. Even though flatness at large distances is energetically favorable, it is counter-balanced by this short-range repulsive effect. Increasing the lattice spacing $l$ and increasing the tension $\sigma$ (that is to say decreasing the repulsion range $\lambda=\sqrt{\kappa/\sigma}$) reduces this short-range repulsion and makes the mixed phase more favorable. This also explains why, in Figure~\ref{rel:stability}, the mixed phase is less unfavorable when $l$ is larger. 

We have also noted that in the mixed case, the large $n$ behavior of $F^{\rm el}(n)$ is linear in $n$, whereas it is super-linear in the pure case. When observing Figure~\ref{rel:stability}, it is likely that, at very large $n$, the mixed phase becomes the more stable again. As discussed in Reference~\cite{Foret08}, this would imply that cluster phases are in fact metastable macro-states rather than stable ones.

\end{document}